\documentclass[preprint, dvipsnames]{elsarticle}

\usepackage{amssymb}
\usepackage{amsmath,amsthm,todonotes,amsfonts}

\theoremstyle{remark}
\newtheorem{remark}{Remark}
\biboptions{sort&compress}

\usepackage[margin=1in]{geometry}

\usepackage[pdftex,
            pdfauthor={Alessandro Alla, Rudy Geelen, Hannah Lu},
            pdftitle={Efficient data-driven regression for reduced-order modeling of spatial pattern formation},
            pdfkeywords={reaction-diffusion models, pattern formation, non-intrusive methods, data-driven modeling, regression}]{hyperref}
\usepackage{url}

\usepackage{algorithm}
\usepackage{algpseudocode}
\usepackage{subcaption}
\usepackage[labelfont=bf,labelsep=period]{caption}
\usepackage[normalem]{ulem}

\usepackage{pgfplots}
\usepackage{xcolor}
\pgfplotsset{compat=1.15}
\usepgflibrary{fpu}

\def \u {{\bf u}}
\def \v {{\bf v}}

\def \f {{\bf f}}
\def \g {{\bf g}}

\journal{Journal of Computational Physics}

\begin{document}

\begin{frontmatter}

\title{Efficient data-driven regression for reduced-order modeling of spatial pattern formation}

\author[label1]{Alessandro Alla} 
\affiliation[label1]{organization={Dipartimento di Matematica, Sapienza Universita di Roma},
            addressline={Piazzale Aldo Moro, 5}, 
            city={Rome},
            postcode={00185}, 
            state={Italy},
            country={}}

            \author[label2]{Rudy Geelen}
\affiliation[label2]{organization={Department of Aerospace Engineering, Texas A\&M University},
            addressline={710 Ross St}, 
            city={College Station},
            postcode={77840}, 
            state={Texas},
            country={USA}}

\author[label3]{Hannah Lu}
\affiliation[label3]{organization={Oden Institute for Computational Engineering and Sciences, The University of Texas at Austin},
            addressline={201 E 24th St}, 
            city={Austin},
            postcode={78712}, 
            state={Texas},
            country={USA}}

\begin{abstract}
We present an efficient data-driven regression approach for constructing reduced-order models (ROMs) of reaction–diffusion systems exhibiting pattern formation. The ROMs are learned non-intrusively from available training data of physically accurate numerical simulations. The method can be applied to general nonlinear systems through the use of polynomial model form, while not requiring knowledge of the underlying physical model, governing equations, or numerical solvers. The process of learning ROMs is posed as a low-cost least-squares problem in a reduced-order subspace identified via Proper Orthogonal Decomposition (POD). Numerical experiments on classical pattern-forming systems—including the Schnakenberg and Mimura--Tsujikawa models—demonstrate that higher-order surrogate models significantly improve prediction accuracy while maintaining low computational cost. The proposed method provides a flexible, non-intrusive model reduction framework, well suited for the analysis of complex spatio-temporal pattern formation phenomena.
\end{abstract}

\begin{highlights}
\item A fully non-intrusive, data-driven reduced-order modeling framework for reaction–diffusion systems with pattern formation.

\item Demonstrated improved prediction accuracy and efficiency over dynamic mode decomposition and operator inference methods for pattern formation problems.

\item Validated on challenging reaction–diffusion benchmarks, showing robust performance and substantial computational speedup.

\end{highlights}


\end{frontmatter}

\section{Introduction}

The classical paradigm in projection-based model reduction involves projecting a high-dimensional dynamical system onto a low-dimensional subspace spanned by a carefully selected set of orthonormal basis, typically obtained through methods such as POD or Krylov subspace techniques~\cite{quarteroni2014reduced,benner2015survey}. This process is \emph{intrusive} in that it requires direct access to the numerical operators of the simulation model that governs the physical system. With particular regard to the modeling of spatial pattern formation, it was found that increasing the reduced space dimension can significantly worsen the model predictions with projection-based methods~\cite{AMS23,AMS24}. This drawback is addressed here by pursuing an approach that avoids explicit projection of the governing equations and, instead, learns a model directly from data provided by high-fidelity simulators~\cite{schmid2010dynamic, tu2013dynamic, kutz2016dynamic, peherstorfer2016data, doi:10.2514/1.J057791, QIAN2020132401, kramer2024learning, 10.1063/5.0170105, chen2021physics,churchill2023flow}. These model reduction methods may thus be classified as \emph{non-intrusive}. In this work, we introduce a methodology that operates in an \emph{equation-free} setting, wherein the governing equations of the physical system are presumed unknown, while retaining the ability to represent nonlinear dynamics via polynomial structure. The approach was found to be particularly suited for application in pattern formation analysis while at the same time lowering the barrier to the broad adoption of these techniques in legacy and commercially available software tools.

Despite their success in various applications, non-intrusive data-driven methods face significant challenges when applied to highly nonlinear dynamical systems. For example, reaction–diffusion systems exhibiting pattern formation~\cite{murray2003mathematical} display rich spatio-temporal behaviors driven by the interplay between local reaction kinetics and diffusion. These systems are well known for generating self-organized spatial patterns—including spots, stripes, labyrinths, and traveling waves—that arise e.g.\ from Turing instabilities under certain conditions~\cite{murray2003mathematical}. Capturing the dynamics of such systems is particularly challenging due to the strong nonlinear coupling of species and spatial modes, which gives rise to dynamic transitions, merging, splitting, and competition among the various patterns. Approximation errors of small magnitude in the reduced-order model can quickly amplify, thereby altering the predictions of the pattern formation dynamics over prolonged timescales. In \cite{AMS23} Turing patterns were investigated within a projection-based model reduction framework. The authors showed that a straightforward POD approach does not yield a monotonically decreasing error when increasing the dimension of the reduced space. To address this issue, they proposed adding a correction term based on the high-fidelity model which compensates for missing information in the reduced-order dynamics. Furthermore, an adaptive strategy was introduced, combined with hyper-reduction techniques to enhance computational efficiency. The adaptivity was guided by the physics of the PDE, which exhibits two distinct regimes: (i) an initial phase where the solution departs from the spatially homogeneous state due to small random perturbations and becomes unstable, and (ii) a subsequent phase where the solution stabilizes towards the steady Turing pattern.

Linear ROM techniques, such as Dynamic Mode Decomposition (DMD) \cite{schmid2010dynamic, tu2013dynamic, kutz2016dynamic}, have fundamental limitations in the context of pattern formation and evolution, as they cannot adequately capture the nonlinearities that drive the system as shown in \cite{AMS24}. Reduced models with linear structure often exhibit degraded predictive ability when extrapolated beyond short-term transients~\cite{lu2020prediction}. This challenge is further compounded by the fact that dimensionality reduction may inadvertently discard critical modes essential for sustaining nonlinear interactions, leading to premature decay, distortion, or failure to capture bifurcations in long-term dynamics. Moreover, while general machine learning-based surrogates, such as deep neural networks, offer universal approximation capabilities, they often operate as black-box models that lack physical interpretability and structural guarantees. Without explicitly embedding the dynamical structure of the system, these models are prone to producing physically inconsistent predictions, unstable long-term behavior, and poor generalization outside of the training regime. Additionally, such models typically require vast amounts of high-fidelity training data to achieve satisfactory accuracy, which can be prohibitively expensive to generate for complex dynamical systems.

Efforts to improve data-driven ROM accuracy for nonlinear systems have led to several important advancements in recent years. One line of work extends DMD through the use of nonlinear observables, inspired by Koopman operator theory~\cite{williams2015data}. By lifting the original state variables into a higher-dimensional feature space, these approaches seek a linear representation of the underlying nonlinear dynamics. However, the effectiveness of Koopman-based methods relies heavily on the selection of appropriate observables. Poorly chosen observables can lead to inaccurate models or fail to capture essential nonlinear behavior, and determining suitable observables often remains problem-dependent and non-trivial. The \emph{piecewise} DMD (pDMD, \cite{AMS24}) was also introduced to construct linear ROMs over shorter temporal horizons, determined by a posteriori error estimates. This approach was applied to Turing pattern datasets and delivered accurate results for small time intervals. The technique also demonstrated some extrapolation capabilities over a few additional time steps, particularly when very close to the target pattern, unlike our method, which consistently delivers accurate predictions over significantly longer time horizons.

In parallel, Operator Inference (OpInf) is a non-intrusive ROM approach that defines a polynomial form for the reduced model based on a set of governing equations, and then learns the corresponding reduced-order operators in a Galerkin-projected subspace using regression \cite{peherstorfer2016data, doi:10.2514/1.J057791, QIAN2020132401, kramer2024learning}. A distinguishing characteristic of OpInf is that it bypasses the need for intrusive access to the full-order governing equations by fitting the reduced operators directly from simulated training data. In contrast to DMD, it also allows for nonlinear terms and is thus able to capture a range of nonlinear dynamics. Because the learning steps of OpInf are rooted firmly in projection-based model reduction, the structure of the ROM is strictly imposed by the physical equations underlying the numerical simulations. This caveat reduces the flexibility of OpInf in cases where the system is not amenable to transformation into low-order polynomial form, or such a transformation creates a large number of auxiliary variables that may compromise prediction accuracy.

In this work, we propose an efficient and fully non-intrusive regression-based framework for ROM of nonlinear dynamical systems exhibiting pattern formation. Our method is motivated by the DMDc framework, which augments the standard DMD formulation by incorporating exogenous inputs (interpreted as control signals) to model forced dynamical systems~\cite{proctor2016dynamic}. While DMDc has been widely applied for systems influenced by external forcing, its mathematical structure---separating intrinsic dynamics from external inputs---offers a natural foundation for representing nonlinearities in an expanded feature space. Building on this philosophy, our method augments the snapshot matrix with polynomial observables to capture nonlinear interactions, which can be interpreted as controls in feedback form, namely the control depends on the state. The resulting surrogate model approximates the dynamics via a low-order polynomial system that faithfully captures the formation, transition, and evolution of complex patterns, especially in regimes where classical linear models fail. Compared to OpInf, our approach is more efficient in regression, as it avoids repetitive least-squares estimation with heavy regularization over tensorized nonlinear terms and without the need of lifting transformation. Indeed, our approach  offers greater flexibility in selecting the polynomial order without requiring a priori specification of the dynamical structure, enabling the model to adaptively increase complexity based on data and learn nonlinear dynamics in a scalable and efficient manner. Numerical experiments on the Schnakenberg and Mimura--Tsujikawa reaction–diffusion systems demonstrate that our method achieves high predictive accuracy over long time horizons at a fraction of the computational cost compared to full-order simulation.

Our proposed framework also bears conceptual similarities to the Sparse Identification of Nonlinear Dynamics (SINDy) approach~\cite{brunton2016discovering}, which aims to identify governing equations from data by selecting a sparse set of terms from a predefined library of candidate functions. However, unlike SINDy, our method does not rely on sparsity-promoting techniques or a fixed function library. Instead, we augment the state space with polynomial observables of user-specified order and directly regress the reduced dynamics without enforcing sparsity or symbolic equation discovery. This offers greater flexibility in capturing complex nonlinear behavior while avoiding sensitivity to library selection or regularization parameters. Recently, in \cite{tran2024weak}, the SINDy framework has been extended to latent spaces using weak-form regression for improved robustness, particularly in combination with POD and autoencoder-based dimensionality reduction. Additionally, while Physics-Informed Neural Networks \cite{raissi2019physics} represent another class of approaches that embed physical constraints into neural network training, they are typically more expensive to train and rely more heavily on explicit knowledge of the governing equations. In contrast, our approach offers a lightweight, interpretable, and data-efficient alternative that is particularly well suited for systems exhibiting pattern formation, where the dominant nonlinear mechanisms are often well approximated by low-order polynomials.

The remainder of this paper is structured as follows. Section~\ref{sec:reaction_diffusion_models} introduces the reaction–diffusion models used to generate pattern-forming dynamics and the dataset construction process. Section~\ref{sec:DMDc} reviews the DMDc framework as a cornerstone of our method. The proposed regression-based method, including its mathematical formulation, connections to DMDc, and comparisons with OpInf are presented in Section~\ref{sec:proposed_method}. In Section~\ref{sec:numerical_experiments} we then demonstrate the effectiveness of our approach through numerical experiments on classical pattern-forming systems. Finally, Section~\ref{sec:con} summarizes the key findings and outlines potential directions for future research.

\section{Reaction-diffusion models for pattern formation}
\label{sec:reaction_diffusion_models}

We are interested in learning data-driven ROMs that accurately capture the pattern formation phenomena governed by Reaction-Diffusion PDE (RD--PDE); see, for instance, \cite{Murray03book} and the references therein for a comprehensive discussion. Accordingly, we consider datasets generated by the following RD--PDE system:
\begin{equation}
\label{RDPDE}
\begin{cases}
\begin{aligned}
\dfrac{\partial u}{\partial t} &= d_u \Delta u + f(u,v) +\beta \mbox{div}(h(u)\nabla u); & (x,y) \in \Omega \subset 
\mathbb{R}^2 , &\quad t \in (0,T], \\[0.25em]
\dfrac{\partial v}{\partial t} &= d_v \Delta v + g(u,v),
\end{aligned}
\end{cases}
\end{equation}
with boundary and initial conditions of the form
\begin{equation}
\label{RDPDE2}
\begin{cases}
\begin{aligned}
(\mathbf{n} \nabla u)_{| \partial \Omega} &= b_u(t), \\
 (\mathbf{n} \nabla v)_{| \partial \Omega}  &= b_v(t), \\
u(x,y,0) &= u_0(x,y) \\
v(x,y,0) &= v_0(x,y),
\end{aligned}
\end{cases}
\end{equation}
where $d_u,d_v\in\mathbb{R}^+$ are the diffusion coefficients, and $T>0$ is the final time of integration. The divergence term is commonly referred to as the \emph{chemotaxis} term, with $\beta\in\mathbb{R}$ representing the chemotaxis sensitivity. The nonlinear reaction terms $f, g : \mathbb{R}^2 \rightarrow \mathbb{R}$ account for biological, chemical and other kind of phenomena.  We consider Neumann boundary conditions, where $\mathbf{n}$ denotes the outward unit normal to the boundary $\partial \Omega$ and $b_u(t), b_v(t): [0,T] \rightarrow \mathbb{R} $ are scalar functions specifying the boundary fluxes. In the case of homogeneous Neumann boundary conditions, these functions are identically zero. 

\begin{figure}[tbp]
    \centering
\begin{subfigure}{.2\linewidth}    \includegraphics[width=\linewidth]{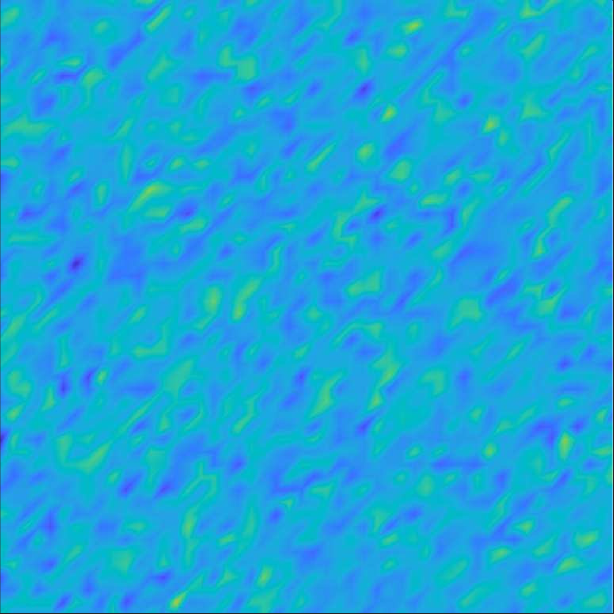}
\caption{$t=0$}
\end{subfigure}
\hspace{0.5em}
\begin{subfigure}{.2\linewidth}    
\includegraphics[width=\linewidth]{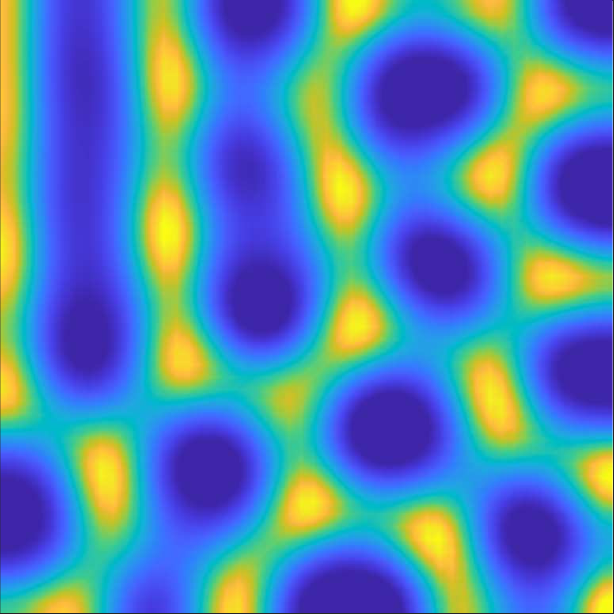}
\caption{$t=33.3$}
\end{subfigure}
\hspace{0.5em}
\begin{subfigure}{.2\linewidth}    \includegraphics[width=\linewidth]{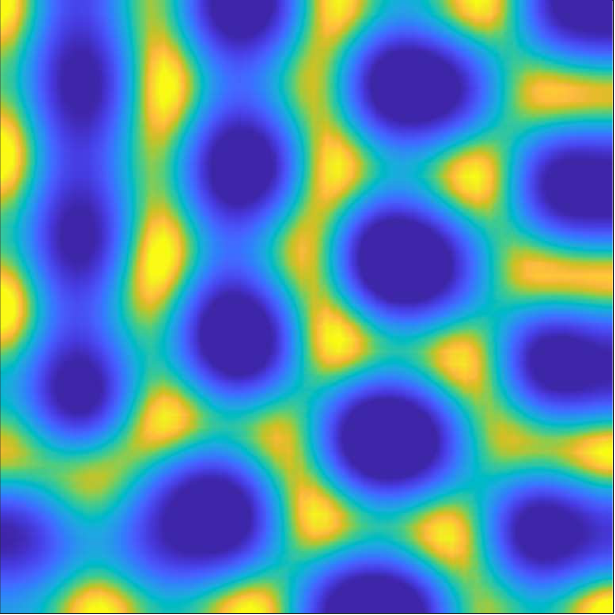}
\caption{$t=100$}
\end{subfigure} \\[0.25em]
\begin{subfigure}{.2\linewidth}    
\includegraphics[width=\linewidth]    {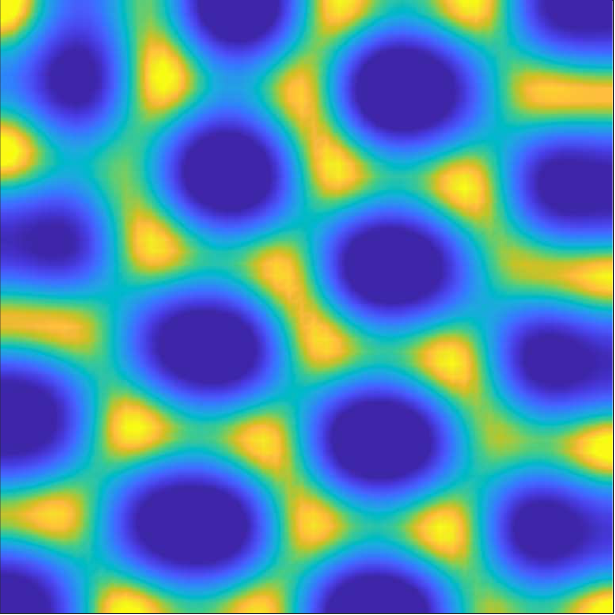}
\caption{$t=133.3$}
\end{subfigure} 
\hspace{0.5em}
\begin{subfigure}{.2\linewidth}    
\includegraphics[width=\linewidth]    {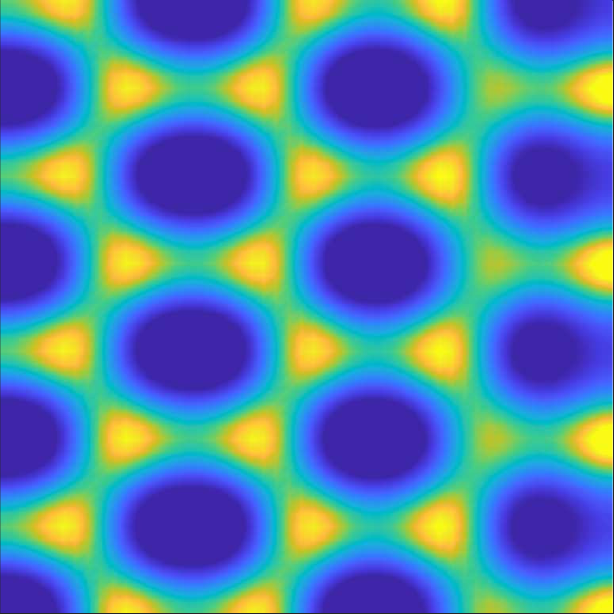}
\caption{$t=200$}
\end{subfigure}    
\hspace{0.5em}
\begin{subfigure}{.2\linewidth}    
\includegraphics[width=\linewidth]    {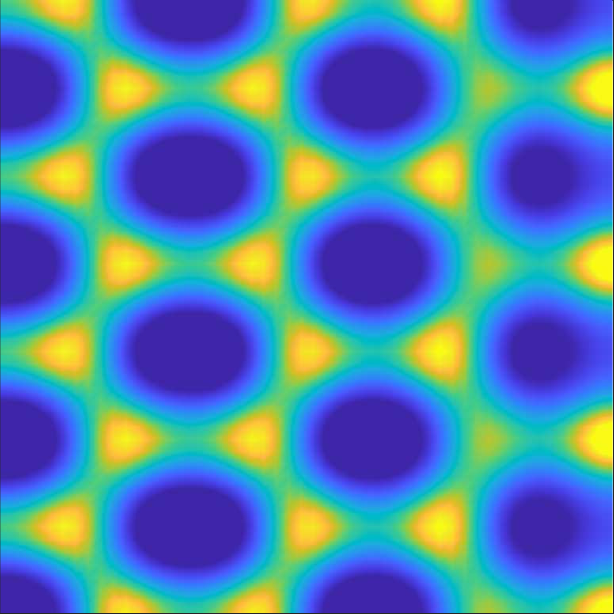}
\caption{$t=500$}
\end{subfigure}      
\caption{Time evolution of hexagonal pattern formation in the Mimura--Tsujikawa model \cite{MT96} at time instances $t=\{0,33.3,100,133.3,200,500\}$ (from top-left to bottom-right).}
\label{fig:snap}
\end{figure}

The PDE model~\eqref{RDPDE}--\eqref{RDPDE2} depends on various parameters that are selected {\it ad hoc} to explore different kinds of dynamics, as detailed below.
When $\beta =0$ and with appropriate choices of configurations for $f,g$ the system exhibits Turing Patterns arising from, for example, (i) the FitzHugh-Nagumo model (ii) the Schnakenberg model~\cite{Madzva03} and (iii) the DIB morphochemical system for battery modeling~\cite{DIB13, DIB15, SLB19}. Conversely, when $\beta\neq0$, the model captures chemotaxis patterns, as observed in, for instance, (i) the Mimura--Tsujikawa model~\cite{MT96} and (ii) the MOMOS models~\cite{HI18}. In Figure~\ref{fig:snap}, we illustrate the pattern formation for the Mimura--Tsujikawa model at different time instances (see Section~\ref{sec:mt} for details about its configuration). The approach to steady state is very slow; for example, the middle and right panels in the top row represent the solution at $t=33.3$ and $t=100$, exhibiting only minor differences. The same slow evolution is evident in the corresponding panels of the bottom row.

This paper uses the method of snapshots from \cite{sirovich1987turbulence} to efficiently compute the dominant modes of dynamical system~\eqref{RDPDE}--\eqref{RDPDE2}. Snapshot data are generated through the repeated numerical simulation of the (usually large-scale) PDE model. Specifically, we apply standard finite differences (\cite{L07}) scheme for the spatial semi-discretization, using a total of $n$ mesh points within the domain $\Omega = [0,L_x] \times [0, L_y]$. The $n$ spatial degrees of freedom arise from discretizing the domain $\Omega$ using $n_x$ and $n_y$ interior mesh points along the $x$ and $y$-directions, respectively. Applying the finite differences with step sizes $h_x = L_x/(n_x+1)$ , $h_y = L_y/(n_y+1)$  yields the following ordinary differential equation (ODE) system:
\begin{equation}
\label{dyn_system}
\begin{cases}
\begin{aligned}
\dfrac{\text{d}{\bf u}} {\text{d}t} &= d_u \mathbf{A} {\bf u}+\f_\beta({\bf u}, {\bf v}), \quad t \in (0,T],\\[0.25em]
\dfrac{\text{d}{\bf v}}{\text{d}t} &= d_v \mathbf{A} {\bf v} + \g({\bf u},{\bf v}), \\[0.25em]
{\bf u}(0) &= {\bf u}_0, \\
{\bf v}(0) &= {\bf v}_0,
\end{aligned}
\end{cases}
\end{equation}
which can be simplified to
\begin{equation}
\label{dyn_system2}
\begin{cases}
\begin{aligned}
\dfrac{\text{d}{\bf w}}{\text{d}t} &= \mathbf F(\mathbf w), \quad t \in (0,T], \\[0.25em]
{\bf w}(0) &= {\bf w}_0,
\end{aligned}
\end{cases}
\end{equation}
where ${\bf w}(t) = ({\bf u}(t), {\bf v}(t))^\top \in \mathbb R^{N_w}$. The unknowns are organized in the standard vectorized form as $\u =\u(t) =(\u_1(t), \dots, \u_{n_x}(t))^\top \in \mathbb{R}^n$, where each block  $\u_i(t)=(u_{i1}, \dots, u_{i,n_y})^\top$ approximates the solution values as  $u_{ij}(t) \approx u(x_i,y_j,t)$ on the given spatial mesh of $n = n_x n_y$ interior points. The same structure is used for the variable $\v$. The compact form $\bf w$ has dimension of $N_w = 2n$. The discrete operator $\mathbf{A} \in \mathbb{R}^{n \times n}$ accounts for the approximation of the Laplace operator $\Delta = \partial_{xx} + \partial_{yy}$ including contributions that enforce homogeneous Neumann boundary conditions (see e.g.\ \cite{DSS20}). The nonlinear term $\f_\beta$ also includes contributions from the divergence term. For further details of the discretization, we refer the reader to~\cite{MDLC25}.

The temporal discretization of~\eqref{dyn_system} is performed using the IMEX Euler scheme, in which the diffusion and nonlinear reaction terms are treated implicitly and explicitly, respectively. Time integration is carried out on the mesh $\tau_{i+1}=\tau_i + h_t$ for $i=0, \dots, n_T$, with a fixed time step $h_t=T/n_T$. To capture the desired patterns, fine spatial resolutions and long integration times ($T \gg 1$) are typically required. As a result, the standard \emph{vector} approach—requiring the solution of a large sparse linear system at each time step—can be computationally expensive. It is worth noting a recent \emph{matrix}-oriented approach proposed in~\cite{DSS20} significantly reduce CPU time in the case $\beta = 0.$

\section{Dynamic mode decomposition with control}\label{sec:DMDc}

The DMDc from \cite{doi:10.1137/15M1013857} serves as the basis for our proposed methodology, to be discussed in Section~\ref{sec:proposed_method}, and is summarized here. The goal of DMDc is to analyze the dynamic relationship of the system variable $\mathbf w(t)\in \mathbb R^{N_w}$ and the control $\mathbf c(t)\in \mathbb R^{N_c}$ from measurement data in the following form:
\begin{equation}\label{eq:DMDc}
\left\{
\begin{aligned}
    \dfrac{\text{d}\mathbf w}{\text{d}t} &= \mathbf A_0 \mathbf c(t) + \mathbf A_1\mathbf w(t), \quad t \in (0,T],\\[0.5em]
    \mathbf w(0) &= \mathbf w_0,
    \end{aligned}
    \right.
\end{equation}
where $\mathbf A_0\in \mathbb R^{N_w\times N_c}$ and $\mathbf A_1\in \mathbb R^{N_w\times N_w}$ are best-fit linear operators learned from measurement data. Let $\mathbf{w}_k \approx \mathbf{w}(t_k)$ denote the measurement at time step $t_k$, and $\Delta \mathbf{w}_k$ denote an approximation of its time derivative. Note that a naive approximation of the derivative from adjacent measurements can produce derivative estimates that are too noisy to be useful, even with noise of moderate amplitude~\cite{van2020numerical}. This calls for more sophisticated methods for data smoothing and/or differentiation of noisy time-series measurements. In what follows, however, it is assumed that the data are (approximately) noise-free. The collection of measurement snapshots can then be arranged into the data matrices
\begin{equation}
\label{eq:data-matrix}
    \mathbf W_\text{R} = \begin{pmatrix}
        |&|&&|\\
        \mathbf w_1&\mathbf w_2&\dots&\mathbf w_{m}\\
          |&|&&|
    \end{pmatrix} \in \mathbb{R}^{N_w\times m}; \quad \mathbf W_\text{L} = \begin{pmatrix}
        |&|&&|\\
       \Delta \mathbf w_1& \Delta\mathbf w_2&\dots& \Delta\mathbf w_{m}\\
          |&|&&|
    \end{pmatrix} \in \mathbb{R}^{N_w\times m}.
\end{equation}
Similarly, the sequence of control input is organized as
\begin{equation}\label{eq:control-matrix}
    \mathbf C = \begin{pmatrix}
                |&|&&|\\
        \mathbf c_1&\mathbf c_2&\dots&\mathbf c_{m}\\
          |&|&&|
    \end{pmatrix} \in \mathbb R^{N_c\times m}.
\end{equation}
Then, problem \eqref{eq:DMDc} can be written in a matrix form:
\begin{equation}
    \mathbf W_\text{L} \approx \mathbf A_0 \mathbf C +\mathbf A_1\mathbf W_\text{R} := \mathbf S\mathbf \Omega,
\end{equation}
where $\mathbf S = \begin{pmatrix}
    \mathbf A_0& \mathbf A_1
\end{pmatrix}\in\mathbb{R}^{N_w\times(N_c+N_w)}$ and $\mathbf \Omega  = \begin{pmatrix}
    \mathbf C &
    \mathbf W_\text{R}
\end{pmatrix}^\top \in\mathbb{R}^{(N_c+N_w)\times m}$. Given $\mathbf \Omega$ and $\mathbf W_\text{L}$, a least-squares solution $\mathbf S$ to the underdetermined problem $\mathbf W_\text{L} = \mathbf S\mathbf \Omega$ can be found by minimizing the Frobenius norm of $\|\mathbf W_\text{L}-\mathbf S\mathbf \Omega\|_F^2$, i.e.,
\begin{equation}
    \mathbf S = \mathbf W_\text{L}\mathbf \Omega^\dagger.
\end{equation}
To solve the above, the reduced Singular Value Decomposition (SVD) is used in $\mathbf \Omega$
\begin{equation}
    \mathbf \Omega \approx \mathbf \Phi \mathbf \Sigma \mathbf \Psi^\top,
\end{equation}
with truncation value $p$. Here $\mathbf \Phi\in \mathbb R^{(N_c + N_w)\times p}$, $\mathbf \Sigma \in \mathbb R^{p\times p}$, and $\mathbf \Psi \in \mathbb R^{m\times p}$. The columns of $\mathbf \Phi$ are orthonormal, so $\mathbf \Phi ^\top \mathbf \Phi = \mathbf I$; similarly, $\mathbf \Psi ^\top\mathbf \Psi = \mathbf I$. 

\begin{equation}
\begin{aligned}
    \begin{pmatrix}
        \mathbf A_0&\mathbf A_1
    \end{pmatrix} = \mathbf S  &= \mathbf W_\text{L}\mathbf \Psi \mathbf \Sigma^{-1} \mathbf \Phi^\top \\
    &= \begin{pmatrix}
        \mathbf W_\text{L}\mathbf \Psi \mathbf \Sigma^{-1} \mathbf \Phi_0^\top&\mathbf W_\text{L}\mathbf \Psi \mathbf \Sigma^{-1} \mathbf \Phi_1^\top
    \end{pmatrix},
\end{aligned}
\end{equation}
where $\mathbf \Phi_0\in \mathbb R^{N_c\times p}$ and $\mathbf \Phi_1\in \mathbb R^{N_w\times p}$.

To find the reduced-order subspace of the output space, a second truncated SVD of rank $r<p$ is needed: 
\begin{equation}
\mathbf W_\text{L} \approx \tilde{\mathbf \Phi}\tilde {\mathbf \Sigma}\tilde{\mathbf \Psi}^\top
\end{equation}
Here $\tilde{\mathbf \Phi}\in \mathbb R^{N_w\times r}$, $\tilde{\mathbf \Sigma} \in \mathbb R^{r\times r}$, and $\tilde{\mathbf \Psi} \in \mathbb R^{m\times r}$. The SVD above is exploited to perform a low-rank truncation of the data. Specifically, if low-dimensional structure is present in the data, the singular values of $\mathbf \Sigma$ will decrease sharply to zero with only a limited number of dominant modes. Using the transformation $\mathbf w \approx \tilde{\mathbf \Phi}\mathbf w^r$, the following reduced-order projection of $\mathbf A_0$ and $\mathbf A_1$ can be computed:
\begin{equation}
\begin{aligned}
    &\mathbf A_0^r = \tilde{\mathbf \Phi}^\top\mathbf A_0   = \tilde{\mathbf \Phi}^\top \mathbf W_\text{L}\mathbf \Psi \mathbf \Sigma^{-1} \mathbf \Phi_0^\top; \\
    &\mathbf A_1^r = \tilde{\mathbf \Phi}^\top\mathbf A_1 \tilde{\mathbf \Phi} = \tilde{\mathbf \Phi}^\top \mathbf W_\text{L}\mathbf \Psi \mathbf \Sigma^{-1} \mathbf \Phi_1^\top\tilde{\mathbf \Phi},
    \end{aligned}
\end{equation}
where $\mathbf A_0^r\in \mathbb R^{r\times N_c}$ and $\mathbf A_1^r\in \mathbb R^{r\times r}$. We can then form the reduced-order equation as follows:
\begin{equation}\label{eq:DMDc-ROM}
\begin{cases}
\begin{aligned}
    \dfrac{\text{d}\mathbf w^r}{\text{d}t} &=\mathbf A_0^r\mathbf c(t) + \mathbf A_1^r\mathbf w^r(t),\quad t\in(0,T],\\[0.25em]
    \mathbf w^r(0) &= \tilde{\mathbf \Phi}^\top  \mathbf w_0.
    \end{aligned}
\end{cases}
\end{equation}

\section{Data-driven polynomial ROMs} \label{sec:proposed_method}

The classical DMD framework approximates the underlying system dynamics by identifying a linear operator that maps between successive snapshot pairs. While extensions such as DMDc, as discussed in Section~\ref{sec:DMDc}, augment the data with exogenous inputs to capture control effects, the same philosophy can be repurposed to capture higher-order nonlinear dynamics in a purely data-driven manner. Motivated by this perspective, we propose an efficient regression-based approach that applies DMD to an augmented snapshot matrix to learn operators for ROMs. Rather than interpreting the augmented terms as control inputs, we systematically expand the snapshot space with higher-order polynomial features that encode nonlinear interactions. This augmentation is particularly crucial for systems whose dynamics are fundamentally nonlinear and cannot be faithfully approximated by a linear ROM. For example, classical DMD fails to capture pattern-forming dynamics~\eqref{RDPDE}--\eqref{RDPDE2} (as presented in~\cite{AMS24,MDLC25}) where the interaction between nonlinear reaction and diffusion terms governs the evolution.

\subsection{Interpretation through the lens of DMDc}
Denote $\mathbf w(t) = [\mathbf u(t),\mathbf v(t)]^\top$ the reference solution to the discretized system~\eqref{dyn_system} using the coupled approach as suggested in \cite{VBGRC22} in a different context. The goal is to find operators $\{\mathbf A_i\}_{i=1}^d$ such that the model ansatz

\begin{equation}
\begin{cases}
    \begin{aligned}
    \dfrac{\text{d}\tilde{\mathbf w}}{\text{d}t} &= \mathbf A_0+\sum_{i = 1}^d\mathbf A_i\left(\tilde{\mathbf w}(t)\right)^{\otimes i},\in t\in(0,T], \\[0.25em]
    \tilde{\mathbf w}(0) &= \mathbf w_0,\\
    \end{aligned}
\end{cases}
\end{equation}
provides accurate predictions $\tilde{\mathbf w}(t)\approx \mathbf w(t)$ for any $t>0$. Here $\left(\tilde{\mathbf w}(t)\right)^{\otimes i}\in \mathbb R^{(N_w)^i}$ contains the components of the $i$-times Kronecker product.

\begin{remark}
The Kronecker product of two vectors $\mathbf{a} \in \mathbb{R}^m$ and $\mathbf{b} \in \mathbb{R}^n$ is defined as
\begin{equation*}
\mathbf{a} \otimes \mathbf{b} = 
\begin{pmatrix}
a_1\mathbf{b} \\
a_2\mathbf{b} \\
\vdots \\
a_m\mathbf{b}
\end{pmatrix}
\in \mathbb{R}^{mn}.
\end{equation*}
We denote $\left(\tilde{\mathbf{w}}(t)\right)^{\otimes i}$ as the $i$-times Kronecker product of $\tilde{\mathbf{w}}(t)$, that is $
\underbrace{\tilde{\mathbf{w}}(t) \otimes \dots \otimes \tilde{\mathbf{w}}(t)}_\text{$i$ terms}$, which contains the components of the Kronecker product up to duplicates due to commutativity of multiplication.
    
\end{remark}

Given a dataset of $m$ snapshots from the reference solution $\{\mathbf w_1,\dots, \mathbf w_{m}\}$, where $\mathbf w_n = \mathbf w(t_n)$ at uniform $h_t$, the problem can be formulated as 
\begin{equation}\label{eq:optim}
\mathbf A_0,\dots, \mathbf A_d = \operatorname*{argmin}_{\hat{\mathbf A}_i\in \mathbb R^{N_w\times (N_w)^i}}\mathcal J_\text{poly}(\hat{\mathbf A}_0,\dots,\hat{\mathbf A}_d),
\end{equation}
with 
\begin{equation}\label{eq:obj}
    \mathcal J_\text{poly}(\hat{\mathbf A}_0,\dots,\hat{\mathbf A}_d) = \sum_{k = 1}^m\|\Delta\mathbf w_{k}-\mathbf{\hat{A}}_0- \sum_{i = 1}^d\hat{\mathbf A}_i\left(\mathbf w_{k}\right)^{\otimes i} \|_2^2,
\end{equation}
and $\Delta \mathbf w_k$  represents the discretized time derivative of $\mathbf w_k$. Follow the same notation as in~\eqref{eq:data-matrix}, equation \eqref{eq:obj} is equivalent~to 
\begin{equation}
      \mathcal J_\text{poly}(\hat{\mathbf A}_0,\ldots,\hat{\mathbf A}_d) = \|\mathbf W_\text{L}-\mathbf {\hat A}_0\cdot \mathbf 1_m^\top-\sum_{i = 1}^d\hat{\mathbf A}_i\left(\mathbf W_\text{R}\right)^{\otimes i} \|_F^2,
\end{equation}
where $\mathbf 1_m\in \mathbb R^{m}$ is a column vector of ones, $\left(\mathbf W_\text{R}\right)^{\otimes i}$ is a  $(N_w)^i$-by-$m$ matrix whose columns are $\left(\mathbf w_{k}\right)^{\otimes i}$.

Due to the high dimensionality of $\{\mathbf A_i\}_{i=1}^d$, the optimization problem~\eqref{eq:optim} is solved on a subspace of order $r \ll N_w$ , i.e., 
\begin{equation}\label{eq:optim_r}
\mathbf A_0^r,\dots, \mathbf A_d^r = \operatorname*{argmin}_{\hat{\mathbf A}_i^r\in \mathbb R^{r\times r^i}}\mathcal J_\text{poly}^r(\hat{\mathbf A}_0^r,\dots,\hat{\mathbf A}_d^r),
\end{equation}
where
\begin{equation}\label{eq:obj_r}
\begin{aligned}
   \mathcal  J_\text{poly}^r(\hat{\mathbf A}_0^r,\dots,\hat{\mathbf A}_d^r) &= \sum_{k = 0}^m\|\Delta \mathbf w^r_{k}-\mathbf{\hat{A}}_0^r- \sum_{i = 1}^d\hat{\mathbf A}_i^r\left(\mathbf w^r_{k}\right)^{\otimes i} \|_2^2, \quad \mathbf w^r_k := \mathbf \Phi^\top\mathbf w_k\in \mathbb R^{r},\\
    & = \|\mathbf W^r_\text{L}-\hat{\mathbf A}_0^r\cdot \mathbf 1_m^\top -\sum_{i = 1}^d \hat{\mathbf A}_i^r\left(\mathbf W^r_\text{R}\right)^{\otimes i}\|_F^2,\quad \mathbf W^r_\text{L,R} := \mathbf \Phi^\top\mathbf W_\text{L,R}.
    \end{aligned}
\end{equation}
The left singular vectors $\mathbf \Phi\in \mathbb R^{N_w\times r}$ are POD modes computed via truncated SVD of rank $r$ of
\begin{equation}\label{eq:svd}
\mathbf W_\text{R} \approx \mathbf \Phi \mathbf \Sigma \mathbf \Psi^\top,
\end{equation}
where $\mathbf \Sigma\in \mathbb R^{r\times r}$, and $\mathbf \Psi\in \mathbb R^{m\times r}$. The SVD in~\eqref{eq:svd} is exploited to perform a low-rank truncation of the data. We denote 
\begin{equation}\label{eq:operator-matrix}
\mathbf S^r = [\mathbf A_0^r,\mathbf A_1^r,\dots, \mathbf A_d^r]\in\mathbb{R}^{r \times \left(\sum_{i =0}^d r^i\right)}
\end{equation}
 and 
\begin{equation}\label{eq:augmented-matrix}
\mathbf \Omega^r = \begin{pmatrix}
    \mathbf 1_m^\top&
    \mathbf W^r_\text{R}&
    \left(\mathbf W^r_\text{R}\right)^{\otimes 2}&
    \dots&
    \left(\mathbf W^r_\text{R}\right)^{\otimes d}    
\end{pmatrix}\in \mathbb R^{\left(\sum_{i =0}^d r^i\right)\times m}
\end{equation}
an augmented snapshots matrix with $\mathbf W^r_\text{R} = \mathbf \Phi^\top \mathbf W_\text{R}$, \eqref{eq:optim_r} can be solved by
\begin{equation}\label{eq:ROM-operators}
\begin{aligned}
\mathbf S^r &= \operatorname*{argmin}_{\hat{\mathbf S}^r\in \mathbb R^{r \times \left(\sum_{i =0}^d r^i\right)}}\|\mathbf W^r_\text{L}-\hat{\mathbf S}^r\mathbf \Omega^r\|_F^2, \\
&= \mathbf\Phi^\top \mathbf W_\text{L}(\mathbf \Omega^r)^\dagger.
\end{aligned}
\end{equation}
Computing the pseudo-inverse $\mathbf \Omega^\dagger$ becomes increasingly expensive as the order $d$ increases. In our numerical experiments, we focus primarily on $d=2$, motivated by the presence of quadratic terms in the pattern formation dynamics. We also implement the method for $d=3$ to explore whether a ROM with higher-order complexity can lead to improved performance. In practice, the choice of order $d$ is constrained by the rank of $\mathbf W_\text{R}$ to maintain an overdetermined system. 
The pseudoinverse $(\mathbf \Omega^r)^\dagger$ is computed with regularization by applying a cutoff to the singular values: only those above a given threshold are retained. This mitigates the effect of near-zero singular values and improves the robustness of the regression.

The operators $\{\mathbf A_i^r\}_{i=1}^d$, which can be decomposed from the $(1+\sum_{j=0}^{i-1}r^j)$-th to $(\sum_{j=0}^{i}r^j)$-th column(s) of $\mathbf S^r$, define a low-dimensional polynomial model of the dynamical system on POD coordinates:
\begin{equation}\label{eq:ROM}
\begin{cases}
\begin{aligned}
\dfrac{\text{d}\mathbf w^r}{\text{d}t} &= \mathbf A_0^r +\sum_{i = 1}^d \mathbf A_i^r \left(\mathbf w^r(t)\right)^{\otimes i},\\[0.25em]
\mathbf w^r(0) &= \mathbf \Phi^\top\mathbf w_0.
\end{aligned}
\end{cases}
\end{equation}
Equation~\eqref{eq:ROM} generalizes the ROM in Equation~\eqref{eq:DMDc-ROM} by removing the dependence on control inputs and incorporating nonlinear dynamics. Specifically, while \eqref{eq:DMDc-ROM} includes a linear control term $\mathbf A_0^r \mathbf c(t)$ and a linear state term $\mathbf A_1^r \mathbf w^r(t)$, Equation~\eqref{eq:ROM} omits the control input and instead includes a constant bias term $\mathbf A_0^r$ along with higher-order polynomial terms in the state variable. This generalization allows the capture of more complex system dynamics in the absence of control forcing. When $d = 1$, the proposed method reduced to generalized DMD~\cite{lu2021extended} and further simplifies to standard DMD in the absence of $\mathbf A_0$. The key steps of the proposed data-driven polynomial ROM approach are summarized in Algorithm~\ref{alg:1}.

\begin{algorithm}[tbp]
\caption{Building data-driven polynomial ROMs}\label{alg:1}
\begin{algorithmic}
\State \textbf{Input:} Collect snapshots data matrices $\mathbf W_\text{L}$ and $\mathbf W_\text{R}$ via~\eqref{eq:data-matrix};
\State \hspace{1em}\textbf{1:} \ Compute  SVD of $\mathbf W_\text{R}$ with rank truncated $r$ as in~\eqref{eq:svd};
\State \hspace{1em}\textbf{2:} \ Construct augmented data matrix $\mathbf \Omega^r$ from projected data matrix $\mathbf W_\text{R}^r$~\eqref{eq:augmented-matrix};
\State \hspace{1em}\textbf{3:} \ Deriving the reduced-order operator matrix $\mathbf S^r$ from solving~\eqref{eq:ROM-operators};
\State \hspace{1em}\textbf{4:} \ Get operators $\{\mathbf A^r_i\}_{i=1}^d$ from the $(1+\sum_{j=0}^{i-1}r^j)$-th to $(\sum_{j=0}^{i}r^j)$-th column(s) of $\mathbf S^r$ for $i = 0, \dots, d$:
$$
\begin{aligned}
&\mathbf{S}^r = 
\begin{pmatrix}
| & | &  & | &  & | &  &  \\
\mathbf{s}_1^r & \mathbf{s}_2^r & \dots & \mathbf{s}^r_{(1+\sum_{j=0}^{i-1}r^j)} & \dots & \mathbf{s}^r_{(\sum_{j=0}^{i}r^j)} & \dots \\
| & | &  & | &  & | &  &
\end{pmatrix}\\
&\hspace{4cm}
\underbrace{\hspace{2.8cm}}_{\mathbf{A}^r_i}
\end{aligned}
$$

\State \hspace{1em}\textbf{5:} \ Integrate \eqref{eq:ROM} in time and approximate the high-dimensional state $\mathbf w(t)$ by projection.

\State \textbf{Output:} $\tilde{\mathbf w}(t) = \mathbf \Phi\mathbf w^r(t)
$
\end{algorithmic}
\end{algorithm}

\subsection{Comparison with operator inference}
\label{subsec:opinf}

The proposed method follows a DMD template in that the underlying, unforced dynamics can be extracted and specified in an \emph{equation-free} manner. It relies only on high-fidelity snapshots, and thus the underlying equations of motion do not have to be known. From a practical standpoint, however, there are notable similarities with OpInf methods for the non-intrusive learning of ROMs. While OpInf \emph{does} take into account explicit knowledge about the structure of the governing equations, both approaches rely on least-squares regression for learning dynamics models from snapshots of temporal data. We here give a brief summary of the standard OpInf method, as described in \cite{QIAN2020132401}.

In the OpInf framework we exploit knowledge of the discrete structure of the matrix operator $\mathbf{F}$ in \eqref{dyn_system2} to inform the structure of the reduced state-space model. In particular, here we seek to expose low-order \emph{polynomial} structure. Polynomial models encompass a large portion of discretized processes in engineering and science. Furthermore, mathematical models that include non-polynomial terms can often be written in polynomial form by leveraging variable transformations that yield a polynomial system of ODEs or differential-algebraic equations~\cite{doi:10.2514/1.J057791, QIAN2020132401, bychkov2024exact}. Importantly, no approximations are invoked in the transformation process. The same mathematical problem is modeled only with different variables, and thus data processing pipeline. Provided that the dynamical system is amenable to transformation into the commonly used quadratic form, state-space model \eqref{dyn_system2} may be written as 
\begin{equation}
    \dfrac{\text{d}\mathbf{z}}{\text{d}t} = \mathbf{A}_0 + \mathbf{A}_1 \mathbf{z}(t) + \mathbf{A}_2  \left(\mathbf{z}(t)\right)^{\otimes 2},
\end{equation}
where $\mathbf{z}(t)\in \mathbb R^{N_z}$ is the transformed (also \emph{lifted}) state variable which comprises of a number of nonlinear terms depending on the specific PDE model that underlies the dynamics. After introducing the low-dimensional POD approximation $\mathbf{z} \approx \boldsymbol{\Phi} \mathbf{z}^r$ with rank $r$, where $\boldsymbol{\Phi} \in \mathbb{R}^{N_z \times r}$ is obtained via the SVD of a data matrix consisting of snapshots of $\mathbf{z}$, similar as in~\eqref{eq:svd}, and applying a Galerkin projection, the reduced-order model takes the form:
\begin{equation}
    \dfrac{\text{d}\mathbf{z}^r}{\text{d}t} = \mathbf{A}^r_0 + \mathbf{A}^r_1 \mathbf{z}^r(t) + \mathbf{A}^r_2 \left( \mathbf{z}^r(t)\right)^{\otimes 2},
\end{equation}
where $\mathbf{A}^r_0 = \mathbf{\Phi}^\top \mathbf{A}_0; \mathbf{A}^r_1 = \mathbf{\Phi}^\top \mathbf{A}_1 \mathbf{\Phi};  $ and $\mathbf{A}^r_2 = \mathbf{\Phi}^\top \mathbf{A}_2 (\mathbf{\Phi} \otimes \mathbf{\Phi})$. While the high-dimensional operators $\mathbf{A}_0, \mathbf{A}_1, \mathbf{A}_2$ are assumed to be known, gaining intrusive access to the source code is often problematic or even prohibited. This restriction can be overcome by learning the reduced-order matrix operators through minimization of the objective function
\begin{equation}
    \mathcal{J}_\text{OpInf}^r(\hat{\mathbf A}^r_0,\hat{\mathbf A}^r_1,\hat{\mathbf A}^r_2) = \sum_{k=1}^m \|     \Delta \mathbf{z}^r_{k} - \hat{\mathbf{A}}^r_0 - \hat{\mathbf{A}}^r_1 \hat{\mathbf{z}}^r_k - \hat{\mathbf{A}}^r_2 \left( \mathbf{z}^r_k \right)^{\otimes 2}
  \|_2^2,
\end{equation}
as in the optimization problem
\begin{equation}\label{eq:optim_opinf}
\mathbf A^r_0,\mathbf A^r_1,\mathbf A^r_2 = \operatorname*{argmin}_{\hat{\mathbf A}^r_i\in \mathbb R^{r\times (r)^i}} \mathcal{J}_\text{OpInf}^r(\hat{\mathbf A}^r_0,\hat{\mathbf A}^r_1,\hat{\mathbf A}^r_2).
\end{equation}
The optimization problem stated in \eqref{eq:optim_opinf} is a linear least-squares problem that can be solved efficiently, as discussed in detail in \cite{kramer2024learning}. Generally speaking, OpInf uses physics-informed Frobenius regularization to promote models with improved predictive capabilities that generalize well over a large range of operating conditions~\cite{mcquarrie2021data, SAWANT2023115836}.

\subsection{Motivating example}
To demonstrate the potential benefits of the proposed polynomial ROM approach presented in Section~\ref{sec:proposed_method} over polynomial models that originate from the OpInf blueprint, consider the one-component reaction-diffusion equation:
\begin{equation}
\label{eq:zfk}
 \dfrac{\partial u}{\partial t} = \dfrac{\partial ^2 u}{\partial x^2}+ R(u); \quad R(u) = \alpha_1 u (1-u) \exp(-\alpha_2(1-u)),
\end{equation}
known as the Zeldovich--Frank--Kamenetskii (ZFK) equation which arises in combustion theory \cite{zeldovich1980flame, doi:10.1137/24M166200X}. We focus on analyzing the change in density profile over a one-dimensional spatio-temporal domain $x \in [-1,1]$ and $t \in [0,1]$ for a parameterized initial condition,
\begin{equation}
    u_0 (x) = 0.5 + 0.5(\mu x + (1-\mu)\sin(-1.5\pi x)),
\end{equation}
where $\mu \in [0,1]$ is a parameter drawn from a distribution to generate different initial conditions. The constants in the model equation are chosen as $\alpha_1 = \alpha_2^2/2$, $\alpha_2 = 10$. 

The methodology proposed in this paper considers the case in which the model has approximate polynomial structure. This choice can be justified by inspecting the governing equation~\eqref{eq:zfk}, which can be written through a Taylor expansion in the form
\begin{equation}
     \dfrac{\partial u}{\partial t} = \dfrac{\partial ^2 u}{\partial x^2} + (\alpha_1 u - \alpha_1 u^2) \exp(-\alpha_2) \left( 1 + \alpha_2 u + \dfrac{\alpha_2^2 u^2}{2} + \mathcal{O}(u^3) \right),
\end{equation}
which is an equation that is polynomial in the dependent variable $u$.

In what follows we will show that the choice of whether or not to use lifting transformations, as discussed in Section~\ref{subsec:opinf}, can have far-reaching implications in terms of prediction accuracy when using polynomial regression. We propose the following lifting map for the ZFK equation:
\begin{equation}
u \mapsto
\begin{pmatrix}
u \\
\exp(-\alpha_2(1-u)) \\
u \exp(-\alpha_2(1- u)) \\
u^2 \exp(-\alpha_2(1- u))
\end{pmatrix}
:=
\begin{pmatrix}
z_1 \\
z_2 \\
z_3 \\
z_4
\end{pmatrix}
\label{eq:zfk_lifting_map}
\end{equation}
The lifted system is then given by
\begin{equation}
\begin{cases}
\begin{aligned}
\dfrac{\partial z_1}{\partial t} &= \dfrac{\partial^2 z_1}{\partial x^2} + \alpha_1 z_3 - \alpha_1 z_4 \\[0.25em]
\dfrac{\partial z_2}{\partial t} &= \alpha_2 z_2 \dfrac{\partial z_1}{\partial t} \\[0.25em]
&= \alpha_2 \dfrac{\partial^2 z_1}{\partial x^2}z_2 + \alpha_1 \alpha_2 z_2 z_3 - \alpha_1 \alpha_2 z_2 z_4 \\[0.25em]
\dfrac{\partial z_3}{\partial t} &= (z_2 + \alpha_2 z_3) \dfrac{\partial z_1}{\partial t} \\[0.25em]
&= \dfrac{\partial^2 z_1}{\partial x^2} ( z_2 +\alpha_2 z_3 ) + \alpha_1 z_2 z_3 - \alpha_1 z_2 z_4 + \alpha_1 \alpha_2 z_3^2 - \alpha_1 \alpha_2 z_3 z_4 \\[0.25em]
\dfrac{\partial z_4}{\partial t} &= (2z_3 + \alpha_2 z_4) \dfrac{\partial z_1}{\partial t} \\[0.25em]
&= \dfrac{\partial^2 z_1}{\partial x^2}(2z_3 + \alpha_2 z_4) + 2\alpha_1 z_3^2 + \alpha_1 \alpha_2 z_3 z_4 - 2\alpha_1 z_3z_4 - \alpha_1 \alpha_2 z_4^2
\end{aligned}
\end{cases},
\label{eq:zfk_lifted_system}
\end{equation}
which is a system of nonlinear PDEs that is clearly quadratic in the lifted variables $z_1, z_2, z_3$ and $z_4$. While other transformations are possible as they are generally non-unique, \eqref{eq:zfk_lifting_map} automatically scales the training data to the $[0,1]$ range, provided that $\beta \geq 0$ and $u$ takes value in $[0,1]$.

\begin{figure}[tbp]
    \centering
    \begin{subfigure}{.57\linewidth}
    \includegraphics[width=\linewidth]{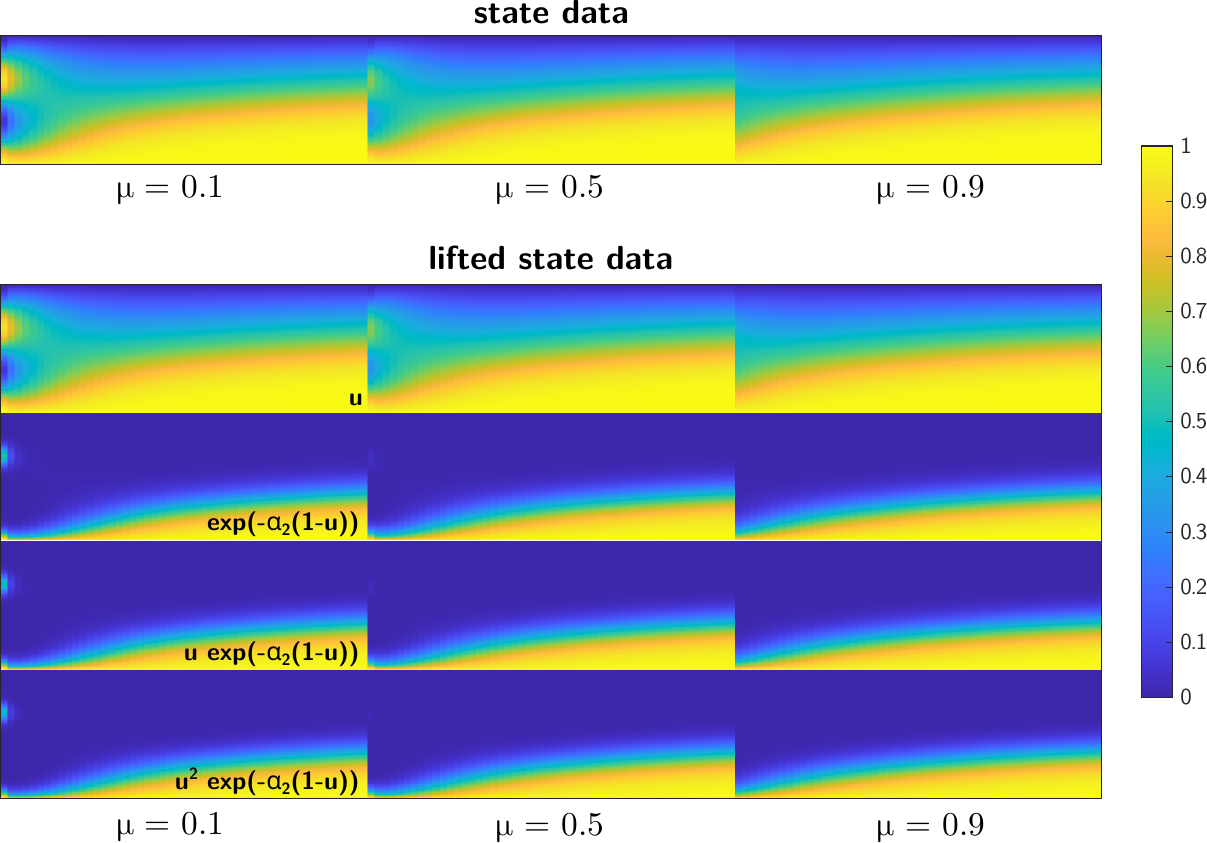}
    \caption{Training data for learning the ZFK equation.}
    \label{fig:zfk_nonlifted}
    \end{subfigure} \hfill
    \begin{subfigure}{.4\linewidth}
       \includegraphics[width=\linewidth]{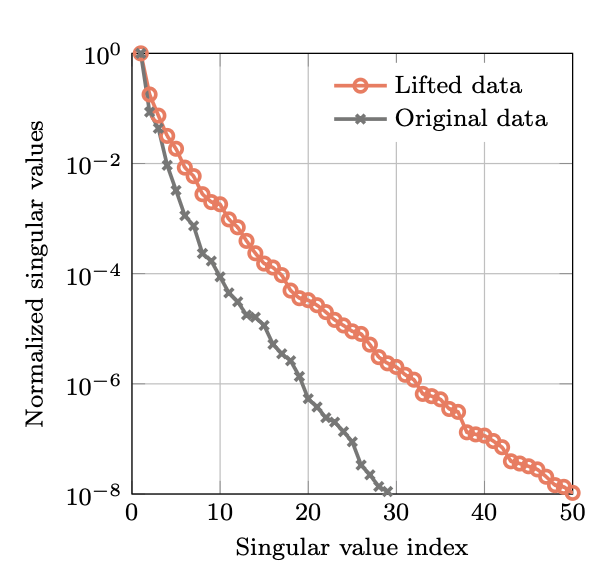}
    \caption{Singular value decay of the snapshot data}
    \label{fig:zfk_lifted}
    \end{subfigure}
    \caption{(a) Training data in original and lifted state-space and (b) the corresponding singular value decay for the ZFK equation \eqref{eq:zfk}.}
    \label{fig:training_data_zfk}
\end{figure}

For learning the low-dimensional models, we start by collecting training data for parameter values $\mu_\text{train} = [0.1,0.5,0.9]$. The data are recorded every 0.02 time units, at each parameter instance, for a total of $m=150$ snapshots. A fourth-order finite difference scheme on adjacent state measurements is used for approximating the time derivatives. The lifting map is applied to the state data to obtain lifted data for $z_1$, $z_2$, $z_3$ and $z_4$. The training datasets and their corresponding singular decay are shown in Figure~\ref{fig:training_data_zfk}. Note that the lifting map increases the dimensionality of the state-space in OpInf models by a factor of four. To test the ability of the ROMs to make predictions about the system under unseen initial conditions, we create a test set in which another 20 trajectories are generated by drawing $\mu$ uniformly on $[0,1]$. Once the models are evaluated on the unseen initial condition data, we compute the error relative to the solution of the original full model defined as $\| \mathbf{W}_\text{test} - \mathbf{W}_\text{pred} \|_F / \| \mathbf{W}_\text{test} \|_F$, where $\mathbf{W}_\text{test}$ is a data matrix containing the unseen test data and $\mathbf{W}_\text{pred}$ the predicted data. While in the proposed method this error metric can be computed directly, OpInf methods that rely on lifting transformations call for a slightly different treatment: we first solve the ROMs to generate predictions, after which the full lifted state is reconstructed from the reduced trajectory, and the lifting transformation is \emph{reversed} to obtain state-space predictions~\cite{QIAN2020132401}.

\begin{figure}[tbp]
    \centering
    \small
    \includegraphics[width=0.4\linewidth]{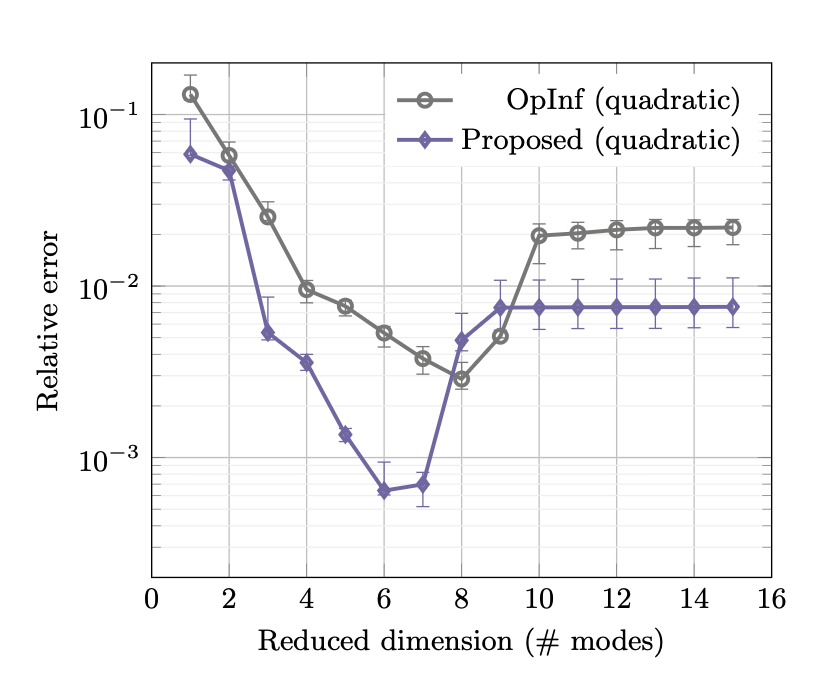}
    \caption{Relative state error in function of the reduced dimension of the ROMs for (a) the OpInf and (b) proposed methods. Both models have linear-quadratic structure. The error bars show the median and first/third quartile errors.}
    \label{fig:zfk_error}
\end{figure}

The test errors are shown in Figure~\ref{fig:zfk_error}. The decrease in the test error with increasing reduced dimension demonstrates the ability of the models created by both approaches to generalize to new inputs. Despite both the proposed and OpInf methods producing ROMs with linear-quadratic structure, the former was found to be more accurate. This observation can be attributed, to certain extent, to the OpInf regression procedure taking place in a coefficient space that balances the independent variable, $u=z_1$, alongside the dependent ones, $z_2, z_3, z_4$. While OpInf ROMs inherit the (here quadratic) structure of lifted system~\eqref{eq:zfk_lifted_system}, they fall short in terms of accuracy even when adding POD modes. The conceptual difference from OpInf is that in the proposed approach model structure is not explicitly preserved. The regression is carried out directly in the coefficient space of the independent variable (here concentration). Even though our ROM methodology therefore no longer has strong ties to projection-based approximation methods, it is the preferred option for this particular model problem.

\section{Numerical experiments}\label{sec:numerical_experiments}

In this section, we provide two additional tests that exhibit pattern formation. In the first, we study the Schnakenberg \cite{Madzva03} model, which accounts for Turing patterns ($\beta = 0$ in \eqref{RDPDE}), while in the second we consider the Mimura--Tsujikawa model \cite{MT96}, which generates hexagonal spatial patterns. We will provide a computational study of the proposed method, comparing ROMs constructed with polynomial degrees $d=\{1,2,3\}$. In each numerical experiment, we show the target pattern from the numerical approximation of the model, along with the initial and final configurations used to train the ROM. Indeed, we do not use the entire dataset from $[0,T]$ for training, for two reasons: (i) to assess the ROM's ability to predict future states, and (ii) because the early dynamics are highly transient and rapidly evolving, making them particularly challenging for ROM training.

Consider a sequence of data samples $\mathbf W = \{\mathbf w_1, \mathbf w_2, \ldots, \mathbf w_m\}$ within the time window $[0,T]$, such that 
\begin{equation}\label{condition}
   \dfrac{\|\mathbf w_{m} - \mathbf w_{1}\|_2}{\|\mathbf w_{1}\|_2} < \varepsilon
\end{equation}
for a prescribed tolerance $\varepsilon > 0$. Note that in this case $\mathbf w_1$ is the first snapshot used for training and does not necessarily coincide with the solution at time $t=0.$ Similarly, $\mathbf w_m$ corresponds to a certain time $t_m<T,$ since we aim to test the extrapolation capabilities of the learned model in the interval $[t_m, T].$

We denote by $\mathbf W^{r,d} = \{\mathbf w_{1}^{r,d}, \ldots, \mathbf w_{{m}}^{r,d},\mathbf w_{{m+1}}^{r,d},\ldots, \mathbf w_{{N_T}}^{r,d}\}$ the solutions of the learned ROM \eqref{eq:ROM}, emphasizing the dimension of the reduced coordinates $r$ and the polynomial degree $d$ in \eqref{eq:ROM}. Note that $\mathbf W^{r,d}$ includes both the results within the training dataset $\{\mathbf w_{1}^{r,d}, \ldots, \mathbf w_{{m}}^{r,d}\}$ and the extrapolation over time $\{\mathbf w_{{m+1}}^{r,d},\ldots, \mathbf w_{{N_T}}^{r,d}\}$ with $\mathbf w_{{N_T}}^{r,d}$ representing the approximate solution of the ROM at the final time $t_{N_T}=T.$

We then study the following relative errors
\begin{align}
\label{err_beh}
    \mathcal{E}_1(r,d) &= \dfrac{\|\mathbf W - \mathbf\Phi\mathbf W^{r,d}(:,1:m)\|_F}{\|\mathbf W\|_F}; \\
\label{err_beh2}
    \mathcal{E}_2(r,d) &= \dfrac{\|\mathbf w_{N_T} - \mathbf w_{N_T}^{r,d}\|_2}{\|\mathbf w_{N_T}\|_2}
\end{align}
to investigate the error behavior within the time frame where the surrogate model has been trained (via $\mathcal{E}_1$) and the extrapolation error at the final time $T$\footnote{We suppose access at the final pattern $\mathbf w_{N_T}$} (via $\mathcal{E}_2$). We also discuss the speed-up factors required to achieve a given level of accuracy, demonstrating that our method is both fast and reliable. We note that the speed-up factor includes the time required to execute Algorithm \ref{alg:1}, covering both the offline and online phases. The time required to generate the training data is not included in this analysis, as it depends on whether such data are already available and on the efficiency of the code used to produce them.

\begin{figure}[tbp]
    \centering
    \begin{subfigure}{0.31\linewidth}
    \centering
    \includegraphics[width=\linewidth]{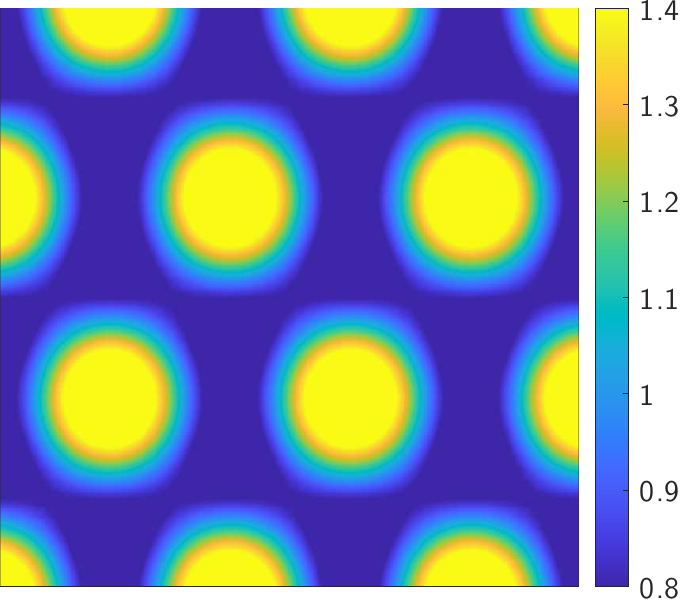}
    \end{subfigure}
    \begin{subfigure}{0.6\linewidth}
    \centering
        \begin{subfigure}{0.4\linewidth}
        \includegraphics[width=\linewidth]{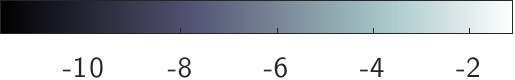}
        \end{subfigure} \\[0.5em]
        \hfill
        \begin{subfigure}{0.45\linewidth}
        \includegraphics[width=\linewidth]{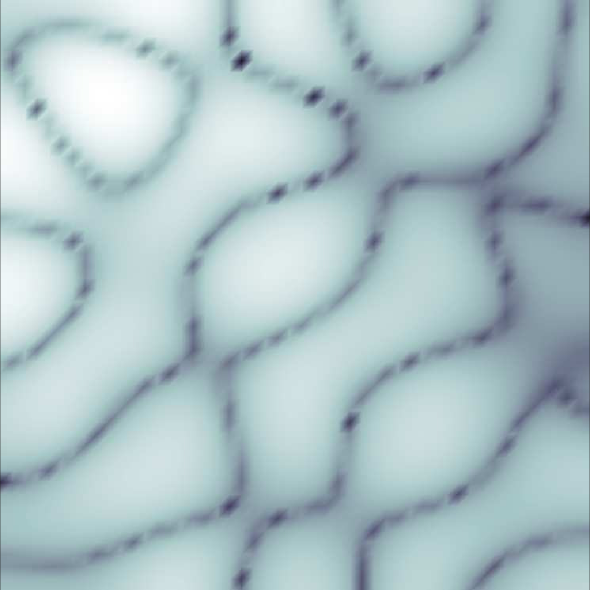}
        \end{subfigure}
        \hfill
        \begin{subfigure}{0.45\linewidth}
        \includegraphics[width=\linewidth]{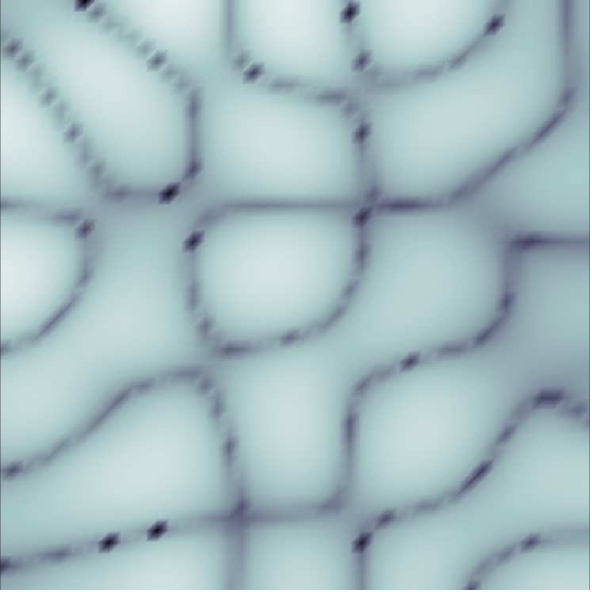}    
        \end{subfigure}
        \hfill
    \end{subfigure}
    \caption{\normalfont{\textsf Test 1}: final pattern $\mathbf u_{N_T}$ (left), absolute differences in log scale $|\mathbf u_m - \mathbf u_1|$ (middle) and $|\mathbf u_m - \mathbf u_{N_T}|$ (right).}
    \label{fig1:pat_diff}
\end{figure}

\subsection{\normalfont{\textsf Test 1}: Schnakenberg model }\label{run1}
We consider the Schnakenberg RD–PDE model with kinetics
\begin{equation}
\label{Schnak_kin}
\begin{cases}
\begin{aligned}
    f(u,v) &= \gamma \left(a - u + u^2 v\right), \\
    g(u,v) &= \gamma \left(b - u^2 v\right),
\end{aligned}
\end{cases}
\end{equation}
which has a unique homogeneous equilibrium at $u_e = a + b$ and $v_e = b/(a+b)^2$ that undergoes a Turing instability.
In RD-PDE~\eqref{RDPDE}--\eqref{RDPDE2} with \eqref{Schnak_kin}, we choose the following parameter values $d_u = 1,\ d_v = 10,\ a = 0.1,\ b = 0.9,\ \gamma = 1000$~\cite{Madzva03}. As initial conditions, we use small random perturbations around $(u_e, v_e)$:
\begin{equation*}
\begin{aligned}
u_0(x,y) &= u_e + 10^{-5} \, \texttt{rand}(x,y),\\ v_0(x,y) &= v_e + 10^{-5} \, \texttt{rand}(x,y),
\end{aligned}
\end{equation*}
where \texttt{rand} denotes the default Matlab function that generates uniformly distributed random values. The spatial domain $\Omega = [0,1] \times [0,1]$ is discretized with $n_x = n_y = 50$ interior points ($n = n_x n_y = 2500$).
The full model is integrated in time using the IMEX-Euler scheme in matrix-oriented form (see \cite{DSS20}) $h_t = 10^{-4}$, for the solutions $u$ and $v$.

The resulting Turing pattern for $\mathbf u_{N_T}$ at the final time $T=2$ is shown in the left panel of Figure~\ref{fig1:pat_diff}. The training window is for $t\in[0.79,0.9495].$ Thus, $\mathbf W\in\mathbb{R}^{5000\times 318}$ since snapshots were stored every $5h_t$ and $\varepsilon=0.05$ in \eqref{condition}. The absolute differences between the training states confirm a significant evolution of the pattern within the training window, highlighting the need for robust data-driven modeling. In the middle panel of Figure~\ref{fig1:pat_diff}, we show $|\mathbf u_{m} - \mathbf u_{1}|$, illustrating the change within the training interval. The right panel shows the absolute difference in log scale between $\mathbf u_{m}$ and the desired Turing pattern $\mathbf u_{N_T}$.

\begin{figure}[tbp]
    \centering
    \begin{subfigure}{.4\linewidth}
    \includegraphics[height=0.8\linewidth,clip, trim=0 0 600 0]{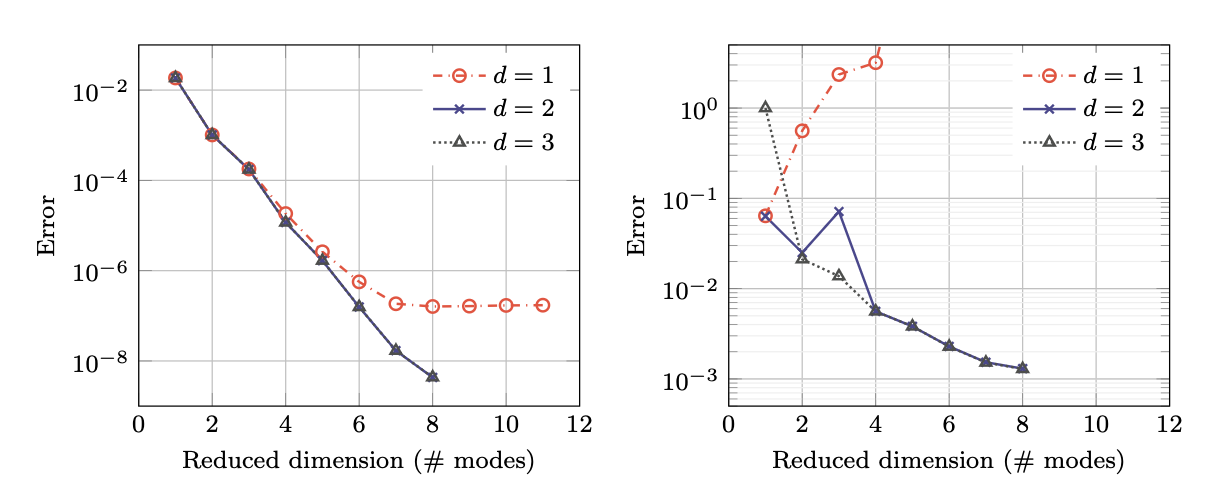}
    \caption{$\mathcal{E}_1(r,d)$ as in \eqref{err_beh}}
    \end{subfigure}
     \begin{subfigure}{.4\linewidth}
       \includegraphics[height=0.8\linewidth,clip, trim=600 0 0 0]{plots_schnack/s1.png}
    \caption{$\mathcal{E}_2(r,d)$ as in \eqref{err_beh2}}
    \end{subfigure}
    \caption{\normalfont{\textsf Test 1}: Relative errors, as defined in \eqref{err_beh}, \eqref{err_beh2}, with respect to the number of POD coordinates.}
    \label{fig1:err}
\end{figure}

In Figure~\ref{fig1:err}, we show the error behavior defined in \eqref{err_beh}, \eqref{err_beh2} for the data-driven models as the number of reduced coordinates, $r$, increases. In the left panel, we report the relative error $\mathcal{E}_1(r,d)$. We observe that all three data-driven models can achieve errors on the order of $10^{-6}$ or lower as $r$ increases. ROMs with $d=\{2,3\}$ were found to be notably more accurate than with $d=1$. In the right panel of Figure~\ref{fig1:err}, we show the behavior of $\mathcal{E}_2$ when predicting the final pattern at time $t=2$. This highlights the extrapolation capability of the ROMs over a prediction window that is larger than the training window. It is clear that our proposed approach yields significantly more accurate results than the linear case, achieving convergence over $r,$ unlike the linear model, which becomes unstable. Nevertheless, although the method is highly reliable, we often find that the quadratic model achieves similar accuracy to the third-order model. 

\begin{remark}
    We note the monotonic decay of the ROM errors with increasing reduced space dimension. In \cite{AMS23} it was shown that achieving such a behavior was not straightforward and called for the inclusion of corrective terms that target the missing information in the reduced models.
\end{remark}

\begin{figure}[tbp]
    \centering
    \scriptsize
    \begin{subfigure}{.32\linewidth}
      \includegraphics[height=0.8\linewidth,clip, trim=0 0 950 0]{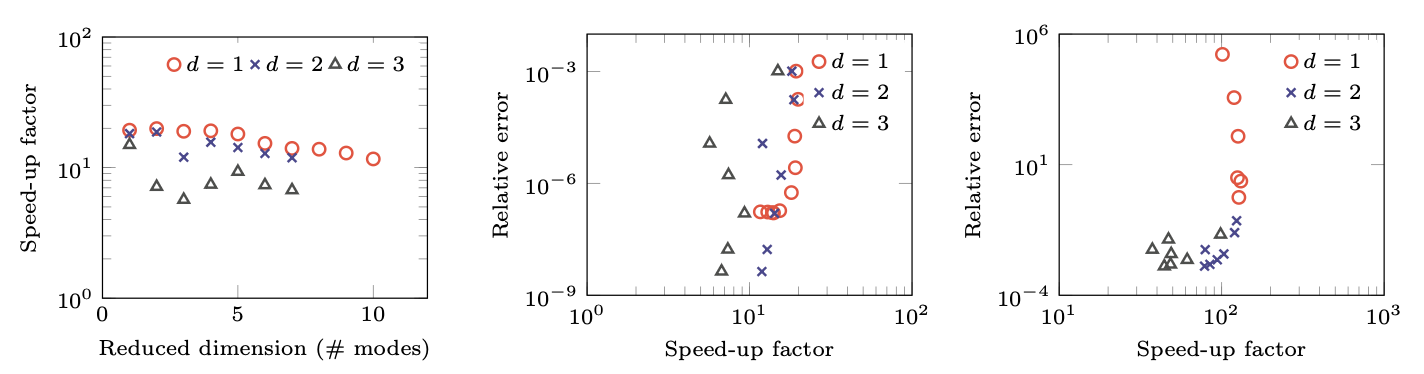}    \caption{}
    \end{subfigure}
    \begin{subfigure}{.32\linewidth}
         \includegraphics[height=0.8\linewidth,clip, trim=480 0 480 0]{plots_schnack/s2.png} 
    \caption{}
    \end{subfigure}
    \begin{subfigure}{.32\linewidth}
            \includegraphics[height=0.8\linewidth,clip, trim=950 0 0 0]{plots_schnack/s2.png} 
    \caption{}
    \end{subfigure}
    \caption{\normalfont{\textsf Test 1}: Speed-up factor over the reduced coordinates (left), Speed-up factor versus $\mathcal{E}_1$ (middle), Speed-up factor versus $\mathcal{E}_2$ (right)} 
    \label{fig1:time}
\end{figure}

Finally, in the left panel of Figure~\ref{fig1:time}, we report the speed-up factor achieved by the three methods. This factor is computed based on the ratio between the CPU time required to build the data-driven models (that is, Algorithm~\ref{alg:1}) and the CPU time of the full-order, high-dimensional simulation model. A more detailed study is documented in the middle and the right panel of the same figure. The middle plot shows the speed-up versus the $\mathcal{E}_1$ error in the training window, illustrating how efficiently the proposed approach reaches a given level of accuracy. As expected, the data-driven model with $d=1$ is faster than $d=2$, which in turn is faster than $d=3.$ Finally, the right panel of Figure~\ref{fig1:time} shows the speed-up versus the extrapolating error $\mathcal{E}_2$. It is visually evident that the linear data-driven model completely fails in this context while the models with $d=2$ and $d=3$ achieve speed-up factors on the order of $\mathcal{O}(10^2)$.
\begin{remark}
    The proposed approach is both fast and reliable. It could also be used as part of a hybrid strategy to accelerate the numerical solution of \eqref{RDPDE}--\eqref{RDPDE2}. For example, one could run a standard solver up to a certain time and then switch to a data-driven method, yielding a substantial speed-up.
\end{remark}

\subsection{\normalfont{\textsf Test 2}: Mimura--Tsujikawa model}
\label{sec:mt}

In the second test, we consider the Mimura--Tsujikawa model, obtained by specifying in \eqref{RDPDE}
\begin{equation}\label{kin_mt}
\begin{cases}
\begin{aligned}
    h(u) &= u \\
    f(u,v) &= q\,u(1 - u) \\
    g(u,v) &= k_1 u - k_2 v
\end{aligned}
\end{cases}
\end{equation}
with parameters $k_1 = 1, k_2 = 32, d_u = 0.0625, d_v = 1, \beta = 17$, and $q = 7$. As initial conditions, we use small random perturbations around the equilibrium $u_e = 1, v_e=k_1/k_2$ as done in \normalfont{\textsf Test 1}. Furthermore, we consider $(x,y) \in \Omega = [0,3] \times [0,3]$ and $T = 500$. The time step is $h_t = 10^{-3}$, and we use the symplectic Euler scheme to solve \eqref{dyn_system} as described in \cite{MDLC25}. The resulting hexagonal pattern $\mathbf u_{N_T}$ at the final time $T$ is shown in the bottom-right panel of Figure~\ref{fig:snap}. Note that data are stored with an effective time step of $h_t = 10^{-2}$, and the training dataset has dimensions $\mathbf W\in\mathbb{R}^{5000\times 703},$ considering $t\in[139.7,209.9]$ and $\varepsilon = 0.1$ in \eqref{condition}. Note again the training window is significantly shorter than the prediction horizon.

\begin{figure}[tbp]
    \centering
    \begin{subfigure}{.4\linewidth}
        \includegraphics[height=0.8\linewidth,clip, trim=0 0 480 0]{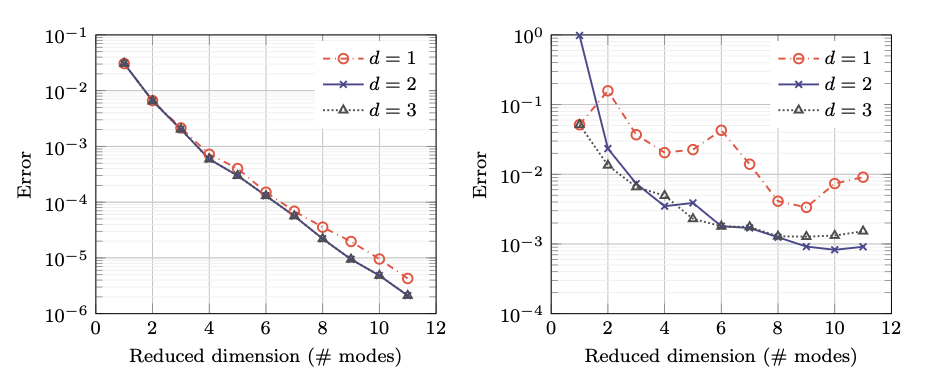}
    \caption{$\mathcal{E}_1(r,d)$ as in \eqref{err_beh}.}
    \end{subfigure}
    \begin{subfigure}{.4\linewidth}
         \includegraphics[height=0.8\linewidth,clip, trim=470 0 0 0]{plots_MIMURA/m1.png}    \caption{$\mathcal{E}_2(r,d)$ as in \eqref{err_beh2}.}
    \end{subfigure}
    \caption{\normalfont{\textsf Test 2}:  Relative errors, as defined in \eqref{err_beh}, \eqref{err_beh2}, with respect the number of POD coordinates.}
    \label{fig2:err}
\end{figure}

\begin{figure}[tbp]
    \centering
    \scriptsize
    \begin{subfigure}{.32\linewidth}
        \includegraphics[height=0.8\linewidth,clip, trim=0 0 720 0]{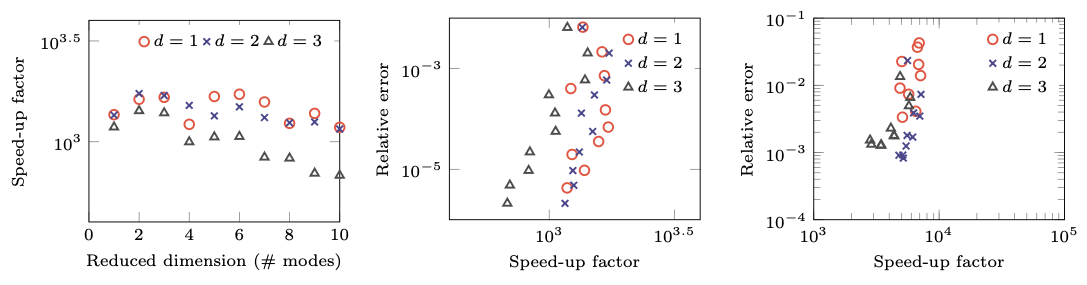}
    \caption{}
    \end{subfigure}
    \begin{subfigure}{.32\linewidth}
          \includegraphics[height=0.8\linewidth,clip, trim=360 0 360 0]{plots_MIMURA/m2.png}    \caption{}
    \end{subfigure}
    \begin{subfigure}{.32\linewidth}
        \includegraphics[height=0.8\linewidth,clip, trim=720 0 0 0]{plots_MIMURA/m2.png}
    \caption{}
    \end{subfigure}   
    \caption{\normalfont{\textsf Test 2}: Speed-up factor over the reduced coordinates (left), speed-up versus $\mathcal{E}_1$ (middle), speed-up versus $\mathcal{E}_2$ (right).}
    \label{fig2:time}
\end{figure}

A quantitative error analysis is presented in Figure~\ref{fig2:err}. The left panel shows the behavior of $\mathcal{E}_1$ as $r$ increases for $d=\{1,2,3\}$. We observe that with only a few modes, for example $r=10$, the error reaches the order of $10^{-6}$ within the training window. However, when using the ROMs outside that time frame, the error increases to the order of $10^{-3}$ for $d=\{2,3\}$. It is clear that our proposed approach provides more accurate extrapolations compared to the linear model. The speed-up factor is shown in the left panel of Figure~\ref{fig2:time}. As expected, increasing the polynomial degree $d$ results in slightly smaller speed-ups. This is also evident when comparing the speed-up needed to reach a certain level of accuracy within the training window (see middle panel of Figure~\ref{fig2:time}). The data-driven model with $d>1$ becomes particularly advantageous in the extrapolation context, where the limitations of the linear approach in terms of accuracy become evident. It is worth noting that, with our approach, we obtain speed-ups of order $\mathcal{O}(10^4)$.

\begin{figure}[tbp]
    \centering
    \begin{subfigure}{.33\linewidth}
    \raisebox{1.5em}{    \includegraphics[width=\linewidth]{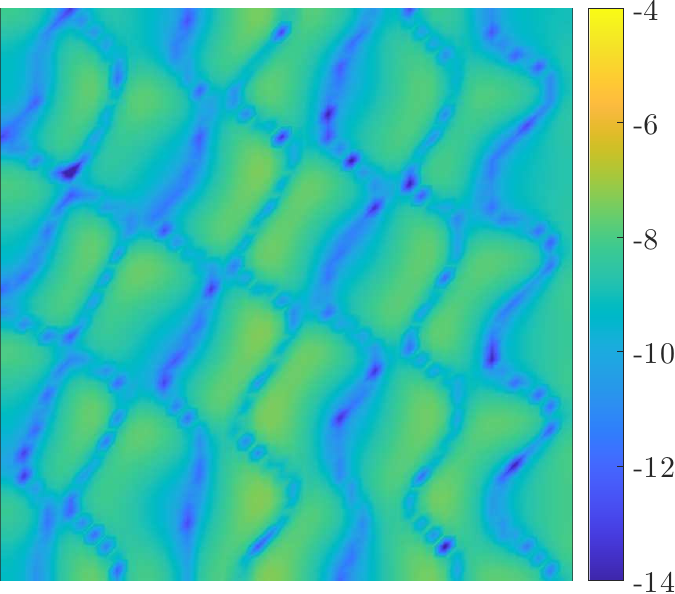}}
    \end{subfigure}
    \hspace{1em}
    \begin{subfigure}{.4\linewidth}
      \includegraphics[width=\linewidth]{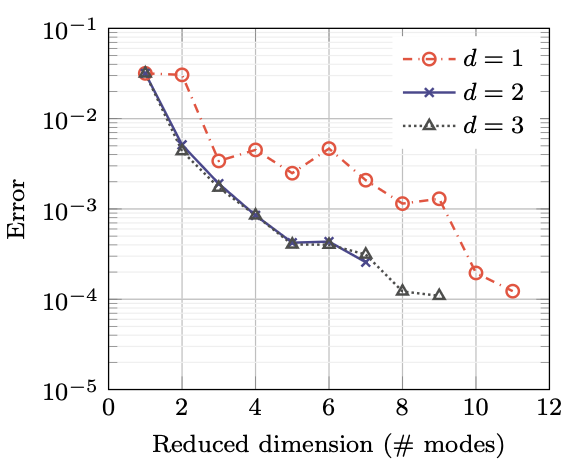} 
    \end{subfigure}
    \caption{\normalfont{\textsf Test 2}: Absolute difference in log scale $|\mathbf u_m-\mathbf u_{N_T}|$ (left), corresponding $\mathcal{E}_2$ error with this new dataset (right).}
    \label{fig22:pat_diff}
\end{figure}

Finally, we present a study for the case $\mathbf W\in\mathbb{R}^{5000\times1098}$ where $\mathbf w_m$ is the approximate solution at time $t=249.9$, maintaining $\varepsilon = 0.1$ in \eqref{condition}. In this scenario, the left panel of Figure~\ref{fig22:pat_diff} which  shows the absolute difference $|\mathbf u_m-\mathbf u_{N_T}|$ in log scale, reveals that the final pattern is already very close to the last data point used to train our ROM. This is also reflected in the extrapolation error shown in the right panel of Figure~\ref{fig22:pat_diff}, where we were able to achieve an error on the order of $10^{-4}$, which is one order of magnitude better than the result shown in Figure~\ref{fig1:err}. This is expected, as having more data closer to the final pattern helps the data-driven models. We observed this behavior consistently across all numerical experiments.

\section{Conclusions}\label{sec:con}

We introduced an efficient, non-intrusive, regression-based framework for constructing ROMs of reaction–diffusion systems that exhibit complex pattern formation. Motivated by the structure of the DMDc framework, our method constructs an augmented observable space by embedding polynomial features of the state variables, thereby enabling the reduced-order model to capture nonlinear interactions in the dynamics. The reduced matrix operators that compose the polynomial ROM are learned via regularized least-squares regression in a low-dimensional space, enabling scalable and accurate ROM construction without requiring access to the underlying high-dimensional equations. 

The proposed approach is entirely data-driven and agnostic to the form of  the underlying equations, making it broadly applicable to any temporally resolved dataset. Its flexibility in part stems from the ability to specify the polynomial order of the observables, which allows the model complexity to be increased systematically based on the nature of the dynamics we are aiming to approximate. We validated our method on two classical reaction–diffusion systems, the Schnakenberg and Mimura–Tsujikawa models, characterized by Turing and chemotaxis driven pattern formation, respectively. Numerical experiments showed that higher-order ROMs (quadratic and cubic) significantly outperform linear models in terms of predictive accuracy. This is especially true in extrapolation scenarios over long periods of time. In addition, the method achieves substantial computational cost reductions, making it a practical and scalable alternative to full-order simulations. Setting all computational considerations aside, the proposed approach offers a flexible and interpretable surrogate modeling framework, distinguishing itself from purely black-box machine learning models.

Building on this foundation, future developments may focus on extending the method to parameter-dependent systems, where the ROM must adapt across regimes of varying parameter sensitivity. Another promising direction involves deriving error estimators to more rigorously assess the model's predictive capabilities and reliability. Finally, addressing scalability through efficient solvers or preconditioning strategies—potentially leveraging recent advances in randomized linear algebra—would further enhance the method's applicability to large-scale, expensive numerical simulations.

\section*{CRediT authorship contribution statement}
The contributions of the authors are equal.
\section*{Data availability}
Data will be made available on request.

\section*{Declaration of competing interest}
The authors declare that they have no known competing financial interests or personal relationships that could have appeared to
influence the work reported in this paper.

\section*{Acknowledgement}
A.\ Alla is member of the INdAM-GNCS activity group. A.\ Alla has developed this work within the activities of the project ``\emph{Data-driven discovery and control of multi-scale interacting artificial agent systems}'' (code P2022JC95T), funded by the European Union – NextGenerationEU, National Recovery and Resilience Plan (PNRR) – Mission 4 component 2, investment 1.1 ``\emph{Fondo per il Programma Nazionale di Ricerca e Progetti di Rilevante Interesse Nazionale}'' (PRIN). A.A. is also supported by MIUR with PRIN project 2022 funds (P2022238YY5, entitled ``\emph{Optimal control problems: analysis, approximation}''). 
 
\bibliographystyle{elsarticle-num}
\bibliography{AGL_draft}

\begin{thebibliography}{10}
\expandafter\ifx\csname url\endcsname\relax
  \def\url#1{\texttt{#1}}\fi
\expandafter\ifx\csname urlprefix\endcsname\relax\def\urlprefix{URL }\fi
\expandafter\ifx\csname href\endcsname\relax
  \def\href#1#2{#2} \def\path#1{#1}\fi

\bibitem{quarteroni2014reduced}
A.~Quarteroni, G.~Rozza, et~al., Reduced order methods for modeling and
  computational reduction, Vol.~9, Springer, 2014.

\bibitem{benner2015survey}
P.~Benner, S.~Gugercin, K.~Willcox, A survey of projection-based model
  reduction methods for parametric dynamical systems, SIAM review 57~(4) (2015)
  483--531.

\bibitem{AMS23}
A.~Alla, A.~Monti, I.~Sgura, {Adaptive POD-DEIM correction for Turing pattern
  approximation in reaction–diffusion PDE systems}, Journal of Numerical
  Mathematics 31~(3) (2023) 205--229.

\bibitem{AMS24}
A.~Alla, A.~Monti, I.~Sgura, {Piecewise DMD for oscillatory and Turing
  spatio-temporal dynamics}, Computers \& Mathematics with Applications 160
  (2024) 108--124.

\bibitem{schmid2010dynamic}
P.~J. Schmid, Dynamic mode decomposition of numerical and experimental data,
  Journal of fluid mechanics 656 (2010) 5--28.

\bibitem{tu2013dynamic}
J.~H. Tu, Dynamic mode decomposition: Theory and applications, Ph.D. thesis,
  Princeton University (2013).

\bibitem{kutz2016dynamic}
J.~N. Kutz, S.~L. Brunton, B.~W. Brunton, J.~L. Proctor, Dynamic mode
  decomposition: data-driven modeling of complex systems, SIAM, 2016.

\bibitem{peherstorfer2016data}
B.~Peherstorfer, K.~Willcox, Data-driven operator inference for nonintrusive
  projection-based model reduction, Computer Methods in Applied Mechanics and
  Engineering 306 (2016) 196--215.

\bibitem{doi:10.2514/1.J057791}
B.~Kramer, K.~E. Willcox, {N}onlinear {M}odel {O}rder {R}eduction via {L}ifting
  {T}ransformations and {P}roper {O}rthogonal {D}ecomposition, AIAA Journal
  57~(6) (2019) 2297--2307.

\bibitem{QIAN2020132401}
E.~Qian, B.~Kramer, B.~Peherstorfer, K.~Willcox, Lift \& {L}earn:
  Physics-informed machine learning for large-scale nonlinear dynamical
  systems, Physica D: Nonlinear Phenomena 406 (2020) 132401.

\bibitem{kramer2024learning}
B.~Kramer, B.~Peherstorfer, K.~E. Willcox, Learning nonlinear reduced models
  from data with operator inference, Annual Review of Fluid Mechanics 56~(1)
  (2024) 521--548.

\bibitem{10.1063/5.0170105}
R.~Geelen, L.~Balzano, S.~Wright, K.~Willcox, Learning physics-based
  reduced-order models from data using nonlinear manifolds, Chaos: An
  Interdisciplinary Journal of Nonlinear Science 34~(3) (2024) 033122.

\bibitem{chen2021physics}
W.~Chen, Q.~Wang, J.~S. Hesthaven, C.~Zhang, Physics-informed machine learning
  for reduced-order modeling of nonlinear problems, Journal of computational
  physics 446 (2021) 110666.

\bibitem{churchill2023flow}
V.~Churchill, D.~Xiu, Flow map learning for unknown dynamical systems:
  Overview, implementation, and benchmarks, Journal of Machine Learning for
  Modeling and Computing 4~(2) (2023).

\bibitem{murray2003mathematical}
J.~D. Murray, Mathematical Biology II: Spatial Models and Biomedical
  Applications, Springer, 2003.

\bibitem{lu2020prediction}
H.~Lu, D.~M. Tartakovsky, Prediction accuracy of dynamic mode decomposition,
  SIAM Journal on Scientific Computing 42~(3) (2020) A1639--A1662.

\bibitem{williams2015data}
M.~O. Williams, I.~G. Kevrekidis, C.~W. Rowley, A data--driven approximation of
  the koopman operator: Extending dynamic mode decomposition, Journal of
  Nonlinear Science 25 (2015) 1307--1346.

\bibitem{proctor2016dynamic}
J.~L. Proctor, S.~L. Brunton, J.~N. Kutz, Dynamic mode decomposition with
  control, SIAM Journal on Applied Dynamical Systems 15~(1) (2016) 142--161.

\bibitem{brunton2016discovering}
S.~L. Brunton, J.~L. Proctor, J.~N. Kutz, Discovering governing equations from
  data by sparse identification of nonlinear dynamical systems, Proceedings of
  the national academy of sciences 113~(15) (2016) 3932--3937.

\bibitem{tran2024weak}
A.~Tran, X.~He, D.~A. Messenger, Y.~Choi, D.~M. Bortz, Weak-form latent space
  dynamics identification, Computer Methods in Applied Mechanics and
  Engineering 427 (2024) 116998.

\bibitem{raissi2019physics}
M.~Raissi, P.~Perdikaris, G.~E. Karniadakis, {Physics-informed neural networks:
  A deep learning framework for solving forward and inverse problems involving
  nonlinear partial differential equations}, Journal of Computational physics
  378 (2019) 686--707.

\bibitem{Murray03book}
J.~Murray, Mathematical Biology II - Spatial Models and Biomedical Applications
  $\{$Interdisciplinary Applied Mathematics V. 18$\}$, Springer-Verlag, Berlin
  Heidelberg, 2003.

\bibitem{MT96}
M.~Mimura, T.~Tsujikawa, Aggregating pattern dynamics in a chemotaxis model
  including growth, Physica A: Statistical Mechanics and its Applications
  230~(3) (1996) 499--543.

\bibitem{Madzva03}
A.~Madzvamuse, A.~J. Wathen, P.~K. Maini, A moving grid finite element method
  applied to a model biological pattern generator, Journal of Computational
  Physics 190~(2) (2003) 478--500.

\bibitem{DIB13}
B.~Bozzini, D.~Lacitignola, I.~Sgura, Spatio-temporal organization in alloy
  electrodeposition: a morphochemical mathematical model and its experimental
  validation, Journal of Solid State Electrochemistry 17~(2) (2013) 467--479.

\bibitem{DIB15}
D.~Lacitignola, B.~Bozzini, I.~Sgura, Spatio-temporal organization in a
  morphochemical electrodeposition model: Hopf and {T}uring instabilities and
  their interplay, European Journal of Applied Mathematics 26~(2) (2015)
  143--173.

\bibitem{SLB19}
I.~Sgura, A.~Lawless, B.~Bozzini, Parameter estimation for a morphochemical
  reaction-diffusion model of electrochemical pattern formation, Inverse Probl.
  Sci. Eng. 27 (2019) 618--647.

\bibitem{HI18}
A.~Hammoudi, O.~Iosifescu, Mathematical analysis of a chemotaxis-type model of
  soil carbon dynamic, Chinese Annals of Mathematics, Series B 39 (2018).

\bibitem{sirovich1987turbulence}
L.~Sirovich, {Turbulence and the dynamics of coherent structures. I. Coherent
  structures}, Quarterly of applied mathematics 45~(3) (1987) 561--571.

\bibitem{L07}
R.~J. Leveque, Finite Difference Methods for Ordinary and Partial Differential
  Equations, Steady State and Time Dependent Problems, SIAM, 2007.

\bibitem{DSS20}
M.~D'Autilia, I.~Sgura, V.~Simoncini, Matrix-oriented discretization methods
  for reaction-diffusion {PDE}s: Comparisons and applications, Comput. Math.
  Appl. 79 (2020) 2067--2085.

\bibitem{MDLC25}
A.~Monti, F.~Diele, D.~Lacitignola, C.~Marangi, Patterns in soil organic carbon
  dynamics: Integrating microbial activity, chemotaxis and data-driven
  approaches, Mathematics and Computers in Simulation 234 (2025) 86--101.

\bibitem{doi:10.1137/15M1013857}
J.~L. Proctor, S.~L. Brunton, J.~N. Kutz, Dynamic mode decomposition with
  control, SIAM Journal on Applied Dynamical Systems 15~(1) (2016) 142--161.

\bibitem{van2020numerical}
F.~Van~Breugel, J.~N. Kutz, B.~W. Brunton, Numerical differentiation of noisy
  data: A unifying multi-objective optimization framework, IEEE Access 8 (2020)
  196865--196877.

\bibitem{VBGRC22}
A.~Viguerie, G.~F. Barros, M.~Grave, A.~Reali, A.~L. Coutinho, Coupled and
  uncoupled dynamic mode decomposition in multi-compartmental systems with
  applications to epidemiological and additive manufacturing problems, Computer
  Methods in Applied Mechanics and Engineering 391 (2022) 114600.

\bibitem{lu2021extended}
H.~Lu, D.~M. Tartakovsky, Extended dynamic mode decomposition for inhomogeneous
  problems, Journal of Computational Physics 444 (2021) 110550.

\bibitem{bychkov2024exact}
A.~Bychkov, O.~Issan, G.~Pogudin, B.~Kramer, Exact and optimal quadratization
  of nonlinear finite-dimensional nonautonomous dynamical systems, SIAM Journal
  on Applied Dynamical Systems 23~(1) (2024) 982--1016.

\bibitem{mcquarrie2021data}
S.~A. McQuarrie, C.~Huang, K.~E. Willcox, Data-driven reduced-order models via
  regularised operator inference for a single-injector combustion process,
  Journal of the Royal Society of New Zealand 51~(2) (2021) 194--211.

\bibitem{SAWANT2023115836}
N.~Sawant, B.~Kramer, B.~Peherstorfer, Physics-informed regularization and
  structure preservation for learning stable reduced models from data with
  operator inference, Computer Methods in Applied Mechanics and Engineering 404
  (2023) 115836.

\bibitem{zeldovich1980flame}
Y.~Zeldovich, Flame propagation in a substance reacting at initial temperature,
  Combustion and Flame 39~(3) (1980) 219--224.

\bibitem{doi:10.1137/24M166200X}
S.~Jelbart, K.~U. Kristiansen, P.~Szmolyan, Traveling waves and exponential
  nonlinearities in the zeldovich–frank-kamenetskii equation, SIAM Journal on
  Applied Dynamical Systems 24~(1) (2025) 530--556.

\end{thebibliography}

\end{document}